%% file: mscol.tex
\documentclass[12pt,preprint]{aastex}

\input epsf.tex

\newcommand\iso[2]{$^{\rm #1}$#2}

\def\teff{\mbox{T$_{\rm eff}$}}
\def\logg{\mbox{log $g$}}
\def\vt{\mbox{v$_{\rm t}$}}
\def\BmV0{\mbox{(B-V)$^{\rm o}$}}
\def\VmK0{\mbox{(V-K)$^{\rm o}$}}
\def\MV0{\mbox{M$_{\rm V}^{\rm o}$}}

\def\carbiso{\mbox{${\rm ^{12}C/^{13}C}$}}
\def\etal{\mbox{et al.}}
\def\eg{\mbox{e.g.}}

\def\third{{3$^{\rm rd}$}}
\def\deg{{$^{\circ}$}}
\def\bd17{\mbox{BD +17\deg 3248}}
\def\cs22{\mbox{CS 22892-052}}

\begin{document}

\title{ 
The Extremely Metal-Poor, Neutron-Capture-Rich Star \cs22: A Comprehensive 
Abundance Analysis\footnote{
Based on observations made at four facilities: 
(a) the NASA/ESA {\it Hubble Space Telescope}, obtained at the
Space Telescope Science Institute (STScI), which is operated by the 
Association of Universities for Research in Astronomy, Inc., under NASA 
contract NAS 5-26555; (b) the Keck~I Telescope of the W. M. Keck Observatory, 
which is operated by the California Association for Research In Astronomy 
(CARA, Inc.) on behalf of the University of California and the California 
Institute of Technology; (c) the H. J. Smith Telescope of McDonald 
Observatory, which is operated by The University of Texas at Austin;
and (d) the Very Large Telescope of the European Southern Observatory 
at Paranal, Chile, from UVES commissioning data and program 165.N-0276(A).
}
}

\author{
Christopher Sneden\altaffilmark{2}, 
John J. Cowan\altaffilmark{2,3},
James E. Lawler\altaffilmark{4},
Inese I. Ivans\altaffilmark{2,5,6},
Scott Burles\altaffilmark{7}, \\
Timothy C. Beers\altaffilmark{8},
Francesca Primas\altaffilmark{9},
Vanessa Hill\altaffilmark{10},
James W. Truran\altaffilmark{11}, \\
George M. Fuller\altaffilmark{12}
Bernd Pfeiffer\altaffilmark{13},
and
Karl-Ludwig Kratz\altaffilmark{13}
}

\altaffiltext{2}{Department of Astronomy and McDonald Observatory, 
University of Texas, Austin, TX 78712; chris@verdi.as.utexas.edu}

\altaffiltext{3}{Department of Physics and Astronomy,
University of Oklahoma, Norman, OK 73019; cowan@mail.nhn.ou.edu}

\altaffiltext{4}{Department of Physics, University of Wisconsin,
Madison, WI 53706; jelawler@facstaff.wisc.edu}

\altaffiltext{5}{Present address: Department of Astronomy, California 
Institute of Technology, Pasadena, CA 91125; iii@astro.caltech.edu}

\altaffiltext{6}{Hubble Fellow}

\altaffiltext{7}{Department of Physics, Massachusetts Institute of Technology, 
77 Massachusetts Avenue, Room 6-113, Cambridge, MA 02139-4307; burles@mit.edu}

\altaffiltext{8}{Department of Physics and Astronomy, Michigan State
University, East Lansing, MI 48824; beers@pa.msu.edu}

\altaffiltext{9}{European Southern Observatory, Karl-Schwarzschild Strasse 2, 
D-85748 Garching bei Muenchen; fprimas@eso.org}

\altaffiltext{10}{GEPI, Observatoire de Meudon, 5 pl. J. Janssen, 
F-92195 Meudon Cedex; Vanessa.Hill@obspm.fr}

\altaffiltext{11}{Department of Astronomy and Astrophysics, Enrico Fermi 
Institute, University of Chicago, 933 E. 56th Street, Chicago, IL 60637;
truran@nova.uchicago.edu} 

\altaffiltext{12}{Department of Physics, University of California at
San Diego, La Jolla, CA 92093-0319; gfuller@ucsd.edu}

\altaffiltext{13}{Institut f\"ur Kernchemie, Universit\"at Mainz,
Fritz-Strassmann-Weg 2, D-55099 Mainz, Germany;
pfeiffer@mail.kernchemie.uni-mainz.de,
 klkratz@mail.kernchemie.uni-mainz.de}

\begin{abstract}
High-resolution spectra obtained with three ground-based facilities and
the Hubble Space Telescope (HST) have been combined to produce a new
abundance analysis of \cs22, an extremely metal-poor giant with large
relative enhancements of neutron-capture elements.
A revised model stellar atmosphere has been derived with the aid of
a large number of Fe-peak transitions, including both neutral and ionized
species of six elements.
Several elements, including Mo, Lu, Au, Pt and Pb, have been detected 
for the first time in \cs22, and significant upper limits have been placed 
on the abundances of Ga, Ge, Cd, Sn, and U in this star.
In total, abundance measurements or upper limits have been 
determined for 57 elements, far more than previously possible.
New Be and Li detections in \cs22\ indicate that the abundances of both 
these elements are significantly depleted compared to unevolved 
main-sequence turnoff stars of similar metallicity.
Abundance comparisons show an excellent agreement between the heaviest
$n$-capture elements (Z~$\ge$~56) and scaled solar system $r$-process 
abundances, confirming earlier results for \cs22\ and other metal-poor stars.
New theoretical $r$-process calculations also show good agreement
with \cs22\ abundances as well as the solar $r$-process abundance components.
The abundances of lighter elements (40~$\leq$~Z~$\leq$~50), however, deviate
from the same scaled abundance curves that match the heavier elements,
suggesting different synthesis conditions or sites for the low-mass  and 
high-mass  ends of the abundance distribution. 
The detection of Th and the upper limit on the U abundance together imply 
a lower limit of 10.4 Gyr on the age of \cs22, quite consistent with
the Th/Eu age estimate of 12.8~$\pm$~$\simeq$ 3~Gyr. 
An average of several chronometric ratios yields an age 
14.2~$\pm$~$\simeq$ 3~Gyr.

\end{abstract}

\keywords{stars: abundances --- stars: Population II --- Galaxy: halo
--- Galaxy: abundances --- Galaxy: evolution --- 
nuclear reactions, nucleosynthesis, abundances}

\section{INTRODUCTION}

Elemental abundance observations of Galactic halo stars provide fundamental 
knowledge about the history and nature of nucleosynthesis in the Galaxy. 
In particular, abundance determinations of the neutron-capture 
($n$-capture) elements --- those from the ($s$)low and ($r$)apid 
neutron-capture processes --- have been utilized to constrain 
early Galactic nucleosynthesis and chemical evolution.
A number of observational and theoretical studies have, for example, 
established that at the earliest times in the Galaxy the $r$-process was 
primarily responsible for $n$-capture element formation
(Spite \& Spite 1978\nocite{spi78}; Truran 1981\nocite{tru81}; 
Sneden \& Parthasarathy 1983\nocite{sne83};
Sneden \& Pilachowski 1985\nocite{sne85}; Gilroy \etal\ 1988\nocite{gil88}; 
Gratton \& Sneden 1994\nocite{gra94}; McWilliam \etal\ 1995\nocite{mcw95}; 
Cowan \etal\ 1995\nocite{cow95}; Sneden \etal\ 1996\nocite{sne96},
Ryan, Norris \& Beers 1996\nocite{rya96}). 
The onset of the $s$-process occurs at higher metallicities (and later 
Galactic times) with the injection of nucleosynthetic material from 
long-lived low- and intermediate-mass stars into the interstellar medium 
(Busso, Gallino, \& Wasserburg 1999\nocite{bus99}; Burris \etal\ 
2000\nocite{bur00}).
The relatively recent detections of radioactive elements such 
as thorium and uranium in a few $r$-process-enhanced metal-poor stars 
are providing new opportunities to determine the ages of the oldest stars, 
and in the process, set lower limits on the age of the Galaxy
(Sneden \etal\ 1996\nocite{sne96}; Cowan \etal\ 1997\nocite{cow97}, 
2002\nocite{cow02}; Johnson \& Bolte 2001\nocite{joh01}; Cayrel \etal\ 
2001\nocite{cay01}; Hill \etal\ 2002\nocite{hil02}; Schatz \etal\ 
2002\nocite{sch02}).

The halo giant star \cs22\ was discovered in the southern Galactic pole 
survey of Beers, Preston, \& Shectman (1985, 
1992)\nocite{bee85}\nocite{bee92}, and was judged to be extremely 
metal-poor based upon the relative weakness of its \ion{Ca}{2} K~line.  
Basic data for this star are listed in Table~1.
McWilliam \etal\ (1995)\nocite{mcw95} included \cs22\ in a followup 
high-resolution chemical composition study of 33 low metallicity stars.
Inspection of the \cs22\ spectrum revealed unusually strong transitions
of $n$-capture elements, and so Sneden \etal\ 
(1994)\nocite{sne94} subjected it to a separate analysis.
That study confirmed the status of \cs22\ as an extremely metal-poor giant
(\teff~$\approx$ 4725~K, \logg~$\sim$ 1.0, [Fe/H]~$\sim$ --3.1).\footnote{
We adopt the usual spectroscopic notations that
[A/B]~$\equiv$ log$_{\rm 10}$(N$_{\rm A}$/N$_{\rm B}$)$_{\rm star}$~-- 
log$_{\rm 10}$(N$_{\rm A}$/N$_{\rm B}$)$_{\odot}$, and
that log~$\epsilon$(A)~$\equiv$ log$_{\rm 10}$(N$_{\rm A}$/N$_{\rm H}$)~+~12.0,
for elements A and B. 
Also, metallicity will be assumed here to be equivalent to the stellar 
[Fe/H] value.}
The low signal-to-noise (S/N) and moderate spectral resolution of their data
permitted study of only the stronger $n$-capture spectral features of 
(mostly) the rare-earth elements. 
But Sneden \etal\ established that \cs22\ is indeed $n$-capture-rich 
(\eg, [Eu/Fe]~$\approx$ +1.6), and that the relative abundances of 
nine elements in the range 56~$\leq$~Z~$\leq$~66 are consistent 
with a scaled solar-system purely $r$-process distribution.
Norris, Ryan, \& Beers (1997a)\nocite{nor97a} performed an independent
study of this star and found abundances consistent with the Sneden \etal\ 
investigation. 

A more extensive spectroscopic study of \cs22\ with better data (Sneden \etal\
1996\nocite{sne96}) expanded the total number of $n$-capture element abundances
to 20, including a clear detection of the strong \ion{Th}{2} 4019~\AA\ line.
Thorium is radioactive with a half-life of 14.0~Gyr, and the observed
[Th/Eu] abundance ratio combined with an assumed extrapolation of the
solar-system $r$-process abundance distribution out to Th yielded
a simple ``decay age'' of about 15~Gyr.
Unfortunately \cs22\ is also relatively carbon-rich ([C/Fe]~$\approx$ +1.0;
McWilliam \etal\ 1995\nocite{mcw95}).
A study of carbon-rich metal-poor stars without $r$-process abundance
enhancements (Norris, Ryan, \& Beers 1997b)\nocite{nor97b} revealed an 
important problem in using the \ion{Th}{2} 4019~\AA\ line
for chronometry estimates: this is a transition blended with not only
atomic contaminants but also two \iso{13}{CH} lines.
All these contaminants must be properly accounted for in the extraction
of the Th abundance in \cs22.

Most strong $n$-capture element transitions in the spectra of metal-poor
giants occur at wavelengths $\lambda$~$<$ 4300~\AA.
Therefore we obtained a new high resolution, high signal-to-noise (S/N) 
near-UV spectrum of \cs22\ with the Keck~I HIRES.
Preliminary abundances from that spectrum were reported by 
Sneden \etal\ (2000)\nocite{sne00a}.
Spectroscopic data farther in the UV were gathered with the Hubble Space
Telescope Imaging Spectrograph (STIS). 
New visible wavelength spectra have been obtained with the McDonald 
Smith 2.7-m telescope and ``2d-coud\'e'' spectrograph.
Commissioning data with the ESO VLT/UT2 (Kueyen) and the UVES
spectrograph have also been included in this assemblage.  
With these four data sets we have performed a new abundance analysis
of \cs22.
Abundances or significant upper limits are reported for 57 elements in
this star. 
To date this is the largest number of elemental abundances discussed
for any star other than the Sun, rivaled only by the
54 elements detected by Cowley \etal\ (2000)\nocite{cow00} in the Ap
star HD~101065 (Przybylski's star).
Observations are described in \S2, the abundance analysis is
given in \S3, and our interpretation of this large abundance set is
discussed in \S4.

\section{OBSERVATIONS AND REDUCTIONS}

{\em HST STIS UV spectra:}
We obtained high-resolution spectra in the vacuum ultraviolet,
2400~$\leq$ $\lambda$~$\leq$ 3100~\AA, with the Hubble 
Space Telescope Imaging Spectrograph ($STIS$).  
The instrument setup included echelle grating E230M centered
at $\lambda$~= 2707~\AA, entrance aperture 0.2$\arcsec\times$0.06$\arcsec$, 
and the $NUV-MAMA$ detector. 
The resolving power ($R$~$\equiv$~$\lambda/\Delta\lambda$) was
approximately 30,000.
Because \cs22\ is a ($V$~=~13.1) K-giant star, 12 separate HST visits
with five individual integrations each were needed to produce
a final reduced spectrum with S/N~$\sim$~20.
We employed standard HST STIS reduction procedures implemented
in IRAF\footnote{
IRAF is distributed by the National Optical Astronomy Observatories, which
are operated by the Association of Universities for Research in Astronomy,
Inc., under cooperative agreement with the National Science Foundation.}
to combine the 60 integrations into a one-dimensional, flat-fielded, 
wavelength-calibrated spectrum.
The individual spectra from each of the 60 integrations had very low S/N 
values, and experiments with the total reductions showed that a more reliable 
final spectrum could be obtained with background corrections neglected.
We therefore used the ``raw'' STIS spectra, and combined these 60
with a median algorithm that neglected the five highest and five lowest
data points in each spectral pixel.

{\em Keck~I Near-UV spectra:}
We gathered Keck~I $HIRES$ (Vogt et~al. 1994\nocite{Vetal94}) echelle 
spectra in the spectral region 3100~$\leq$ $\lambda$~$\leq$ 4250~\AA.
Seven individual integrations were combined to produce the final spectrum.
The resolving power of the fully reduced spectrum was $R$~$\simeq$ 45,000.
Its S/N decreased blueward, from $\sim$150 per pixel at
$\lambda$~$\sim$ 4200~\AA\ to $\sim$20 at $\lambda$~$\sim$ 3100~\AA;
both the stellar flux and the CCD response decline toward the UV. 
We used auxiliary tungsten lamp spectra for flat-fielding the stellar
spectrum and Th-Ar lamp spectra for wavelength calibration.
We performed all extraction and reduction tasks to produce a spectrum
for analysis with the software package MAKEE 
(e.g., Barlow \& Sargent 1997\nocite{bar97}).

{\em McDonald 2d-coud\'e spectra:} 
We obtained a high-resolution spectrum in the wavelength range
4300~$\leq$ $\lambda$~$\leq$ 9000~\AA\ with the McDonald 
Observatory 2.7m H.~J. Smith telescope and ``2d-coud\'e'' 
echelle spectrograph (Tull et~al. 1995)\nocite{tul95}.
In addition to tungsten and Th-Ar calibration spectra, we also
acquired a spectrum of the hot, rapidly rotating star $\lambda$~Aql
for use in canceling telluric (O$_2$ and H$_2$O) lines that 
occur in the yellow-red region.
The spectrograph setup had a 1.2$\arcsec$ entrance slit and
a Tektronix 2048$\times$2048 CCD detector, yielding a 2-pixel
resolving power of $R$~$\simeq$ 60,000.
The star was observed on three consecutive nights.  
Initial processing of the raw spectra was done with standard 
$IRAF$ echelle reduction tasks.  
The co-added reduced spectra have S/N~$>$~100 for 
$\lambda$~$>$ 5500~\AA, and this declines
steadily to levels of $\sim$40 at $\lambda$~$\sim$ 4400~\AA.

{\em VLT UVES spectra:}
We made use of VLT UVES (Dekker \etal\ 2000\nocite{dekker00}) 
echelle spectra which were obtained during the commissioning phase 
of the instrument\footnote{
Data release of March 2000, discussed in
\mbox{http://www.eso.org/science/uves\_comm/UVES\_comm\_stars.html}}
and during the course of the ESO Large Program 165.N-0276(A) 
({\it First Stars}, PI {\it R. Cayrel}). 
The two sets of observations obtained using a dichroic
beam-splitter covered respectively 3300-4000~\AA\ + 4800-10000~\AA\ (at a
resolution of $R$~$\approx$~55000) and 3350-4570\AA\ + 6700-10000\AA\ 
($R$~$\approx$~45000), for a total of 5.5 hours of integration.
The spectra were reduced individually, using a MIDAS (Warmels 
1991\nocite{war91} and references therein) context dedicated
to UVES, including bias and inter-order background subtraction, optimal
extraction of the object (weighted over the profile object profile),
flat-field correction and a wavelength calibration using a Th-Ar lamp.
The S/N of the combined spectra ranged from $\sim$40 at 3300\AA\ to 
$\sim$100 at 4500\AA\ in the blue part of the spectrum, and 
$\sim$175-200 throughout most of the red part.

{\em Final processing of the spectra:}
We used the software package $SPECTRE$ (Fitzpatrick \& 
Sneden 1987)\nocite{fit87} for the completion of the stellar data 
reductions, to $(a)$ eliminate anomalous radiation events and bad
pixels; $(b)$ normalize spectral continua with spline function fits 
to interactively chosen continuum points; and $(c)$ smooth the spectra 
via 2-pixel FWHM Gaussian convolutions for STIS, McDonald,
and those parts of the Keck spectra with $\lambda$~$<$ 3400~\AA. 
For unblended spectral lines, equivalent widths (EWs) were measured by 
fitting Gaussian profiles or by direct integrations over the absorption 
profiles.
Line data for transitions used in this study are given in Table~2 
for lighter (Z~$\leq$~30) elements, and in Table~3 for $n$-capture elements.
Many lines are blended and/or have multiple sub-components, so synthetic 
spectra described in the next section were used instead of EWs for the
abundance computations.
There are two categories of such complex spectral features.
Some of them (\eg, \ion{Ba}{2} 4554.03~\AA, \ion{La}{2} 4086.71~\AA, or
\ion{Tb}{2} 3702.85~\AA) are sufficiently strong and unblended that 
measurement of an approximate EW was possible.  
For those lines the EW entries are written as ``syn(EW)'' so that
the reader may have an indication of the absorption strengths.
However, these EWs were not and should not be employed in abundance 
computations for \cs22892.
Other synthesized transitions are either so weak or so blended
that EW measures would be meaningless and misleading.
For those lines the EW entries are simply ``syn''.

\section{ABUNDANCE ANALYSIS}

\subsection{Model Atmosphere and Abundances of Lighter Elements}

Spectral lines of Fe-peak species were employed in deriving a new model 
atmosphere for \cs22.
A conventional LTE analysis was performed in the same manner as was done in 
our earlier abundance analyses of $n$-capture-rich, very metal-poor stars 
(\eg, Westin \etal\ 2000\nocite{wes00}; Cowan \etal\ 2002\nocite{cow02}).
We began with the line list of Westin et~al. for
species with Z~$\leq$~30, augmenting it to include lines of additional 
Fe-peak species rarely employed in abundance analyses 
(\eg, \ion{Sc}{1}, \ion{Mn}{2}).
Single-line EW matches were employed for most transitions, but full
spectrum syntheses were done for transitions of \ion{Sc}{1}, \ion{Sc}{2},
\ion{Mn}{1}, \ion{Mn}{2}, and \ion{Cu}{1}, all of which have substantial
hyperfine substructures (hfs).

The effective temperature, \teff, was determined by minimizing the 
line-to-line abundance differences with excitation potential, EP,
for species with a sufficient EP range.
The microturbulent velocity parameter, \vt, was adjusted to remove trends
in line abundances with EW. 
The gravity, log~$g$, was set by requiring neutral and ionized species of 
Sc, Ti, V, Cr, Mn, and Fe to have the same abundances on average.

Trial model atmospheres for \cs22\ were interpolated from the
Kurucz (1995)\nocite{kur95}\footnote{see http://cfaku5.cfa.harvard.edu/}
grid with a modified version of software kindly provided by 
A. McWilliam (private communication).
Models without convective overshooting were chosen, as recent literature
suggests that such models give best overall matches to optical and UV
spectra of metal-poor stars (\eg, Castelli, Gratton, \& Kurucz 
1997\nocite{cas97}; Peterson, Dorman, \& Rood 2001\nocite{pet01}).
Abundances were computed for each line with the current version of Sneden's
(1973)\nocite{sne73} line analysis code, iterating on the model
parameters until the satisfaction of the above criteria was achieved.
The best model had parameters (\teff, \logg, [M/H], \vt)~=
(4800$\pm$75~K, 1.5$\pm$0.3, --2.5, 1.95$\pm$0.15).
The chosen model metallicity, [M/H]~= --2.5, is somewhat larger than the 
derived [Fe/H]~= --3.1, but --2.5 is the lowest metallicity of the Kurucz 
grid models without convective overshooting.
Test abundance runs using models with overshooting and smaller model 
metallicities yielded little differences in derived model parameters.

Our spectroscopically-derived model atmosphere for \cs22\ is in good
accord with low-resolution indicators.
Extensive photometric temperature scales for giant stars have 
been determined by Alonso, Arribas, \& Martinez-Roger 
(1999, 2001)\nocite{aam99}\nocite{aam01}.
We used their formulas with the photometric data of Table~1
to compute \teff\ values for color indices $U-V, B-V, V-K, J-H$, and $J-K$
(the $V-R$ color is not covered by the Alonso formula). 
Giving the derived \teff($V-K$) double weight due to the excellent 
temperature sensitivity of the $V-K$ color, we obtain: 
(a) $<\teff>$~= 4791~$\pm$~18~K ($\sigma$~= 44~K) if $E(B-V)$~= 0.00
(Beers \etal\ 1999)\nocite{bee99};
(b) $<\teff>$~= 4836~$\pm$~24~K ($\sigma$~= 58~K) if $E(B-V)$~= 0.02
(Beers \etal\ 1992)\nocite{bee92},
and (c) $<\teff>$~= 4861~$\pm$~27~K ($\sigma$~= 65~K) if $E(B-V)$~= 0.03
(from dust maps of Schlegel \etal\ 1998)\nocite{sch98}.
All of these estimates agree well with the spectroscopic \teff.

To compute an evolutionary gravity for \cs22, fundamental physical 
considerations provide a standard formula
relating absolute visual magnitude ($M_V$~= --0.2~$\pm$~0.4, 
Beers \etal\ 1999\nocite{bee99}), effective temperature 
(\teff~= 4800~$\pm$~75~K), mass (assumed here to be a metal-poor 
near-turnoff mass $\mathcal{M}$~= 0.75~$\pm$~0.10~$\mathcal{M}_{\odot}$), 
and bolometric correction (BC~= --0.39~$\pm$~0.04, computed from 
Eq.~18 of Alonso \etal\ 1999)\nocite{aam99}:
$$
\log g_{evol} = -12.50 + 0.4\left(M_V + BC\right) +
\log \mathcal{M} + 4\log \teff\
$$ 
These quantities combine to yield \logg$_{evol}$ = 1.85~$\pm$~0.25, in
agreement with our derived \logg$_{spec}$ = 1.5~$\pm$~0.3, given
the uncertainties of both values.

Note that \logg$_{spec}$~$<$~\logg$_{evol}$ is in the expected sense if the
ionization equilibrium used to constrain \logg$_{spec}$ is affected by
``overionization'' departures from LTE.
However, the close agreement of abundances from neutral and ionized
species of six different Fe-peak elements (with ionization potentials
that range from 6.8 to 7.9~eV) indicates that our derived spectroscopic
gravity is sufficient to compute reliable relative  abundances of elements 
with only one ionization state or the other detected in \cs22.
Exceptions may exist for neutral species of elements with very low first
ionization potentials (I.P.~$\lesssim$~5~eV); see remarks for
\ion{K}{1} below.

\begin{figure}
\epsscale{1.0}
\plotone{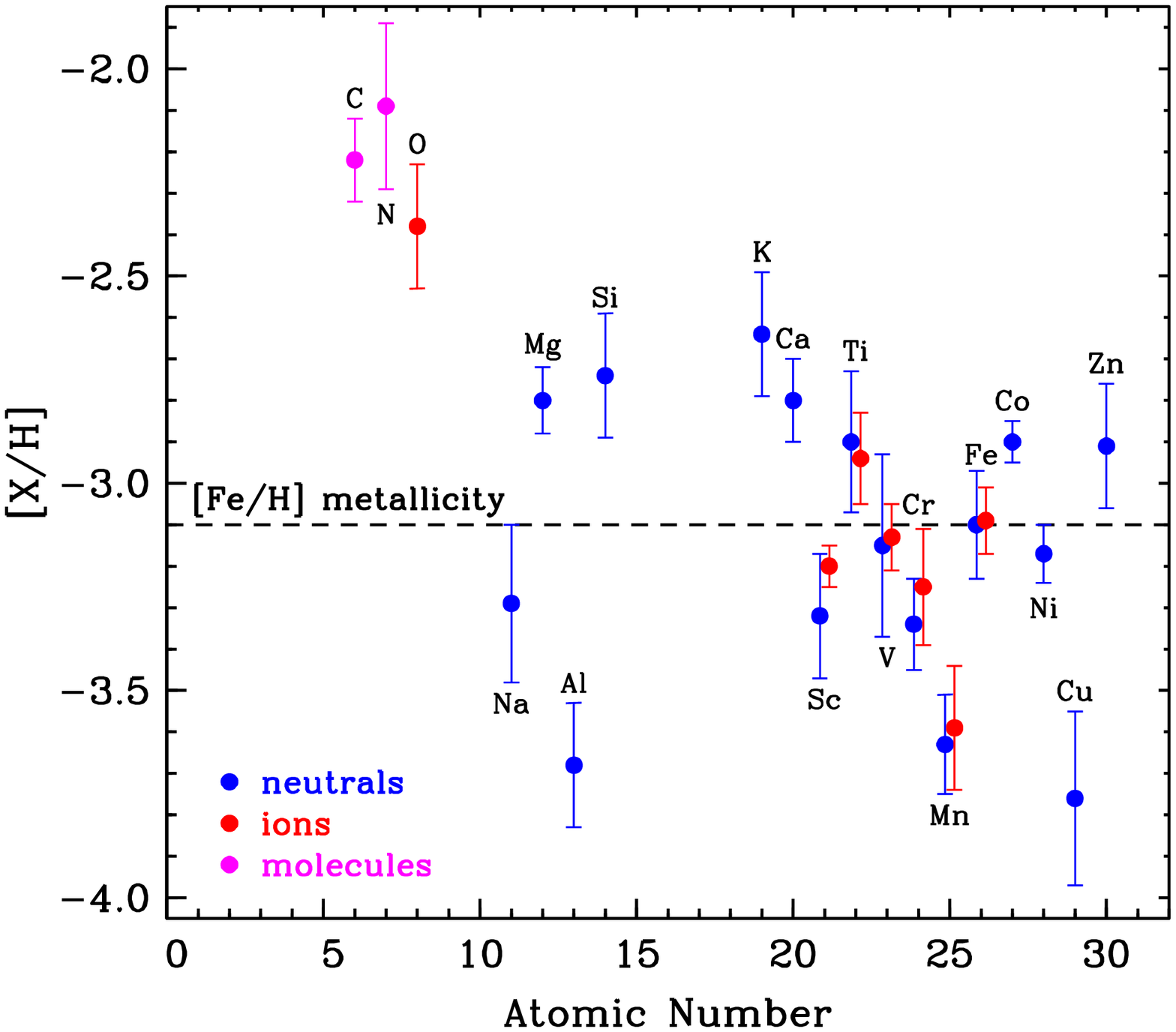}
\caption{
Abundances with respect to the Sun, [X/H], versus atomic number for elements
with Z~$\leq$~30.
The symbols are defined in the figure legend, and the dashed line representing
the standard ``metallicity'' of \cs22\ is drawn through the \ion{Fe}{1}
abundance, [Fe/H]~= --3.10.
See \S 3.1 for discussion of the error bars displayed in this and other
figures.
For display purposes small leftward and rightward shifts for points from 
neutral and ionized species, respectively, have been imposed on abundances 
of elements represented by two ionization stages.
\label{f1}}
\end{figure}

Derived abundances of elements with Z~$\leq$~30 are listed in 
Table~4 and plotted in Figure~\ref{f1}.
Solar abundances are taken from Grevesse \& Sauval (1998)\nocite{gre98}.
Following Cowan \etal\ (2002)\nocite{cow02}, the error bars in this
figure are generally the sample standard deviations $\sigma$.
However, minimum values were imposed, noting that the smallest line-to-line
scatter for any species (of light or $n$-capture elements) with at least 
several measured lines is $\sigma$~$\sim$~0.05~dex.
Therefore for an abundance based on three or more lines, the plotted error
bar is the maximum of its $\sigma$ and 0.05. 
For a species with two measured lines, the error bar is the maximum of 
its $\sigma$ and 0.10.
For a species with only one line, $\sigma$ is undefined and so an 
arbitrary value of 0.15 is adopted as the error bar.
In a few cases with only 1-2 lines for a species, we have increased the 
error bar even further to account for very uncertain $gf$ values, 
difficult line blending problems, etc.

In general the abundances displayed in Figure~\ref{f1} conform
to well-established patterns for these elements in very metal-poor halo stars
(\eg, Cayrel 1996\nocite{cay96}; McWilliam 1998\nocite{mcw98}).
In particular, the $\alpha$ elements are all overabundant in \cs22: 
$<$[Mg,Si,Ca,TiI,TiII/Fe]~= +0.26~$\pm$~0.04 ($\sigma$~= 0.08).
The light odd-Z elements Na and Al are underabundant with respect to Fe.
Among the Fe-peak elements, the now-well-established overabundance of
Co and underabundances of Mn and Cu (perhaps Sc and Cr as well) are evident.
Apparently Zn is also overabundant, as it is in some other
stars of this metallicity (\eg, Primas \etal\ 2000c\nocite{pri00c}; 
Blake \etal\ 2001\nocite{bla01}).

A few of these overall conclusions should be viewed with caution.
We emphasize the inherent uncertainties of elemental abundances deduced
from only one line, as in the cases of Al, Si, and Zn.
Also, the Cu abundance of \cs22\ has been determined from the \ion{Cu}{1}
resonance lines at 3247 and 3274~\AA, not from the 5105~\AA\ line that is
employed in most other studies of this element in metal-poor stars.
Since the 5105~\AA\ line cannot be detected on our \cs22\ spectra,
the mapping of its Cu abundance onto [Cu/Fe] trends with metallicity
(\eg, Mishenina \etal\ 2002\nocite{mis02}) cannot be accomplished with
confidence yet.
The Al abundance is derived from the \ion{Al}{1} 3961~\AA\ resonance line, 
which is known (\eg, Ryan \etal\ 1996\nocite{rya96}) to yield 
systematically lower abundances than do the red-region spectral 
lines arising from excited states of this species.
Finally, the large overabundance of K derived here may or may not be real.
Takeda \etal\ (2002)\nocite{tak02} have made statistical equilibrium 
calculations of the \ion{K}{1} 7699~\AA\ line strengths in stars of various 
metallicities, concluding that LTE-based [K/Fe] values are generally 
too large by 0.2 to 0.7~dex.
Further investigation of this issue is warranted, but it is beyond the
scope of the present investigation.

Next we consider the CNO element group.
The [\ion{O}{1}] 6300.3~\AA\ line is detected on our VLT spectrum, but it is
extremely weak, EW~= 1.9~$\pm$~0.2~m\AA.
The line does not appear to be contaminated by telluric features.
A synthetic spectrum fit to this spectral region was used to estimate
the O abundance, and to ensure that reasonable abundances could be
recovered from the comparably weak neighboring \ion{Fe}{1} 6301.5,
6302.5~\AA\ lines.
The derived abundance, [O/Fe]~$\approx$ +0.7, is in good agreement
with abundances determined from [\ion{O}{1}] lines in other very
low metallicity giant stars (\eg. Westin \etal\ 2000\nocite{wes00};
Depagne \etal\ 2002\nocite{dep02}).
However, the Grevesse \& Sauval (1998) recommended value of
log~$\epsilon_\sun$(O)~= 8.83 has recently been re-assessed by
Reetz (1999)\nocite{ree99}, Allende Prieto, Lambert, \& Asplund 
(2001)\nocite{all01} and Holweger (2001)\nocite{hol01}; the mean of 
their values is log~$\epsilon_\sun$(O)~= 8.75.
Adoption of this solar O value of course would increase the
relative abundance in \cs22\ to [O/Fe]~$\approx$ +0.8.
It is recommended that the [\ion{O}{1}] 6300~\AA\ be re-observed in \cs22\
with better resolution and S/N (the 6363~\AA\ line is a factor of three weaker, 
and likely to remain undetected in this star).

The anomalous strength of the CH G-band in \cs22\ was first discovered from a
medium-resolution spectrum of this star, and noted in Table~8 of Beers \etal\
(1992)\nocite{bee92}. 
The implied large C abundance was confirmed in subsequent high-resolution 
studies (Sneden \etal\ 1996\nocite{sne96}; Norris \etal\ 1997a)\nocite{nor97a}.
Our synthetic spectrum fits to G-band lines recover the earlier results.
But as with O, the solar abundance of C may need revision,
as a new analysis of the [\ion{C}{1}] 8727~\AA\ line by Allende Prieto,
Lambert, \& Asplund (2002)\nocite{all02} yields 
log~$\epsilon_\sun$(C)~= 8.39, about 0.1~dex smaller than the value
recommended by Grevesse \& Sauval (1998)\nocite{gre98} (however, Holweger 
2001\nocite{hol01} recommends log~$\epsilon_\sun$(C)~= 8.59).
This of course would lead to a comparable upward shift in our
derived [C/Fe] = 0.95, but a careful analysis of the CH G-band
in both the Sun and \cs22\ should be done before altering the
[C/Fe] listed in Table~4.

Norris \etal\ (1997b)\nocite{nor97b} searched without success for 
\iso{13}{CH} features on their spectra of \cs22, and suggested that 
\carbiso~$\gtrsim$~10.
Using the Sneden \etal\ (1996)\nocite{sne96} CTIO 4-m echelle spectrum,
Cowan \etal\ (1999)\nocite{cow99} estimated \carbiso~$\approx$~16,
From the new VLT spectrum we concur with the earlier value,
deriving \carbiso~=~15~$\pm$~2.
Additionally, CN blue system $\Delta$v~=~0 bandheads are weakly visible in 
\cs22\ (see Figure~6 of Sneden \etal). 
New syntheses of the 3868-3885~\AA\ region of our Keck HIRES spectrum 
yield approximately the same N enhancement ([N/Fe]~$\approx$ +0.8) as 
was claimed in the earlier work.

\begin{figure}
\epsscale{0.9}
\plotone{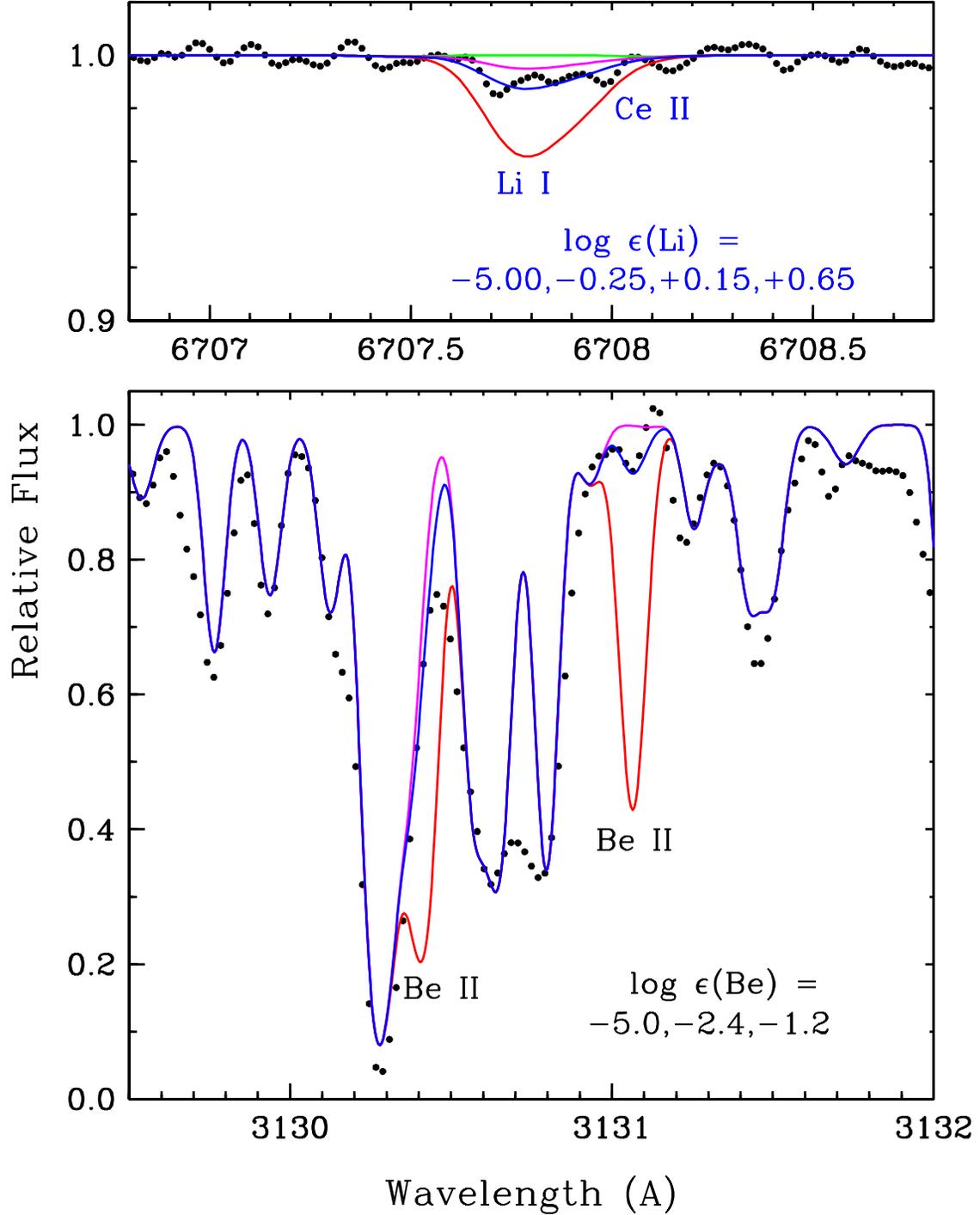}
\caption{
Observed and synthetic spectra of the \ion{Li}{1} and \ion{Be}{2} resonance
doublets in \cs22.
In the upper panel, a potential \ion{Ce}{2} blending transition at
6708.10~\AA\ is marked.
The synthetic spectra are computed with log~$\epsilon$(Li)~= --5.0
(dotted line), --0.25 (short dashed line), +0.15 (solid line), and
+0.65 (long dashed line).
In the lower panel, the synthetic spectra are chosen to illustrate the
appearance of the spectrum without the presence of Be
(log~$\epsilon$(Be)~= --5.0, dotted line), the appearance if Be were 
assumed to be undepleted (log~$\epsilon$(Be)~= --2.4, dashed line), and the 
best-fit to the 3131~\AA\ feature (log~$\epsilon$(Be)~= --1.2, solid line).
See \S 3.1 for further comment on the Li and Be features.
\label{f2}}
\end{figure}

Finally, we comment on Li and Be abundances, listed in Table~4
but not plotted in Figure~\ref{f1}.
These two elements are normally destroyed in stellar interiors during
evolution beyond the main sequence, so their present abundances in \cs22\
are unlikely to reflect the values of the ISM from which this star formed.
The \ion{Li}{1} 6707.8~\AA\ resonance doublet is tentatively detected in
our VLT spectrum, with EW~= 3.5~$\pm$~0.5~m\AA.
Synthetic spectrum fits, using the \ion{Ca}{1} 6717.69~\AA\ line to
align the wavelength scales of the observed and synthesized spectra, and 
with no \iso{6}{Li} included in the syntheses, yielded
log~$\epsilon$(Li)~$\approx$ +0.15.
We show the \ion{Li}{1} doublet spectrum in the upper panel of
Figure~\ref{f2}. 
While it is possible to argue for lower Li abundances
than we have estimated, much larger abundances are ruled out based on
the extreme weakness of the observed feature.
A new study of the 6707.8~\AA\ feature in $s$-process-enriched post-AGB
stars by Reyniers \etal\ (2002)\nocite{rey02} suggests that all or part of
this often-strong absorption may actually be due to a line of \ion{Ce}{2}
($\lambda$~= 6708.099~\AA, E.P.~= 0.71~eV, log~$gf$~= --2.12; Palmeri
\etal\ 2000\nocite{pal00}).
We included this line in our syntheses, but (as shown in Figure~\ref{f2})
it is well separated from the \ion{Li}{1} in the \cs22\ spectrum, and
will create detectable line depth only if the Ce abundance is increased
by factors of 5-10 more than our derived value.
The Li abundance of \cs22\ is not unusual for a metal-poor giant star
that has undergone convective envelope mixing, as several studies
(\eg, Ryan \& Deliyannis 1998\nocite{rya98}, Gratton \etal\
2000\nocite{gra00}) find log~$\epsilon$(Li)~$\lesssim$~0
for stars with similar atmospheric parameters.

The Be abundance in \cs22\ was determined from our Keck spectrum,
using spectrum synthesis fits to the observed \ion{Be}{2} 3130~\AA\ doublet;
see the lower panel of Figure~\ref{f2}.
We used the line list assembled and extensively tested by
Primas \etal\ (1997)\nocite{pri97}, fixing all
abundances except Be to the values given derived in this paper.
There are several $n$-capture species transitions that surround the
\ion{Be}{2} doublet: \ion{Ce}{2} 3130.34~\AA, \ion{Nb}{2} 3130.78~\AA,
\ion{Gd}{2} 3130.81~\AA, \ion{Ce}{2} 3130.87~\AA.
These lines are usually insignificant in normal dwarfs and subgiants,
but do contribute to the total absorption in \cs22.
We derived log~$\epsilon$(Be)~= --2.4~$\pm$~0.4 from the 3131.07~\AA\ line.
The large error estimate given in Table~4 is due to the
combination of substantial line blending, the weakness of the \ion{Be}{2}
doublet, and the modest S/N of our spectrum in this region.

The Be abundance in \cs22\ is extremely low.
Halo dwarf stars of similar metallicity typically have Be abundances 
about a factor of 10 larger.
Two examples are: log~$\epsilon$(Be)~= --1.1~$\pm$~0.2 in both G~64-12 
([Fe/H]~= --3.3; Primas \etal\ 2000a\nocite{pri00a}) and LP~815-43
([Fe/H]~= --2.95; Primas \etal\ 2000b\nocite{pri00b}).
Alternatively, using the Boesgaard \etal\ (1999)\nocite{boe99} mean 
relationship between Be abundance and stellar metallicity,
the predicted initial abundance for \cs22\ with [Fe/H]~= --3.1
would be log~$\epsilon$(Be)~= --1.6.
These zero-age Be values in very metal-poor stars indicate that significant 
internal Be depletion has occurred in \cs22.
For illustration we show in Figure~\ref{f2} a synthetic spectrum
with log~$\epsilon$(Be)~= --1.2; the observed 3131.07~\AA\ and
(highly blended) 3130.42~\AA\ lines both clash badly with this 
``zero-age'' Be abundance.

The low Be value in \cs22, combined with the equally low Li abundance, 
may hint at the presence of a mixing mechanism that is able to affect 
both Li and Be at the same time.
Detailed predictions of this phenomenon at low metallicities are
lacking, but Boesgaard \etal\ (2001) show that depletion of both
elements occurs in metal-rich F-type stars.
Their observations yield a depletion correlation of
$\delta$[log~$\epsilon$(Be)]~$\approx$ 0.36$\delta$[log~$\epsilon$(Li)].
Assuming that \cs22\ began its life with a ``Spite plateau'' Li abundance
of log~$\epsilon$(Li)~$\approx$ +2.0 (Ryan \etal\ 1999\nocite{rya99} 
and references therein) then for \cs22\ the observed 
log~$\epsilon$(Li)~= +0.15 implies
$\delta$[log~$\epsilon$(Li)]~$\sim$ --1.9 or 
$\delta$[log~$\epsilon$(Be)]~$\sim$ --0.7.
This prediction is in rough accord with the observed 
$\delta$[log~$\epsilon$(Be)]~$\sim$ --1.0.

Unfortunately, we cannot confidently predict the Be abundance
that \cs22\ had at birth, because there may be significant star-to-star
scatter among the lowest metallicity unevolved stars.
For example, Primas \etal\ (2000b)\nocite{pri00b} derive 
log~$\epsilon$(Be)~$\leq$ --1.4 for CD~--24\deg17504 ([Fe/H]~= --3.3), 
at least a factor of two lower than that of G~64-12.
This possible initial scatter, combined with a lack of extensive Be studies
in metal-poor giants and lack of quantitative predictions for Be depletion
in low metallicity stars, prevent us from drawing more quantitative
conclusions about the Be abundance in \cs22.

\subsection{Abundances of the $n$-Capture Elements}

Derivation of $n$-capture elements was accomplished with EW matches
for lines that were either very weak (log(EW/$\lambda$)~$\lesssim$ --5.3)
or single $and$ unblended absorptions.
For other transitions synthetic spectrum computations were used.
All $n$-capture lines are listed in Table~3.
Construction of atomic/molecular line lists for the synthetic spectra
have been described in earlier papers (\eg, Sneden \etal\ 1996\nocite{sne96};
Westin \etal\ 2000\nocite{wes00}; Cowan \etal\ 2002\nocite{cow02}).
Laboratory studies in the past couple of decades have provided a wealth 
of accurate transition probabilities, hfs constants, and 
isotopic wavelength shifts for many $n$-capture-element neutral and ionized
transitions detectable in the spectra of \cs22\ and other $r$-process-rich
stars.
A good summary and references to line parameters of lanthanide elements
has been given by Wahlgren (2002)\nocite{wah02}; a similar compendium
for other $n$-capture elements would be welcome.
Our line data are briefly discussed in Appendix~A, with emphasis on species
detected in the present study for the first time in this star.
New laboratory results for a few spectra are included in Appendix B.

\begin{figure}
\epsscale{1.0}
\plotone{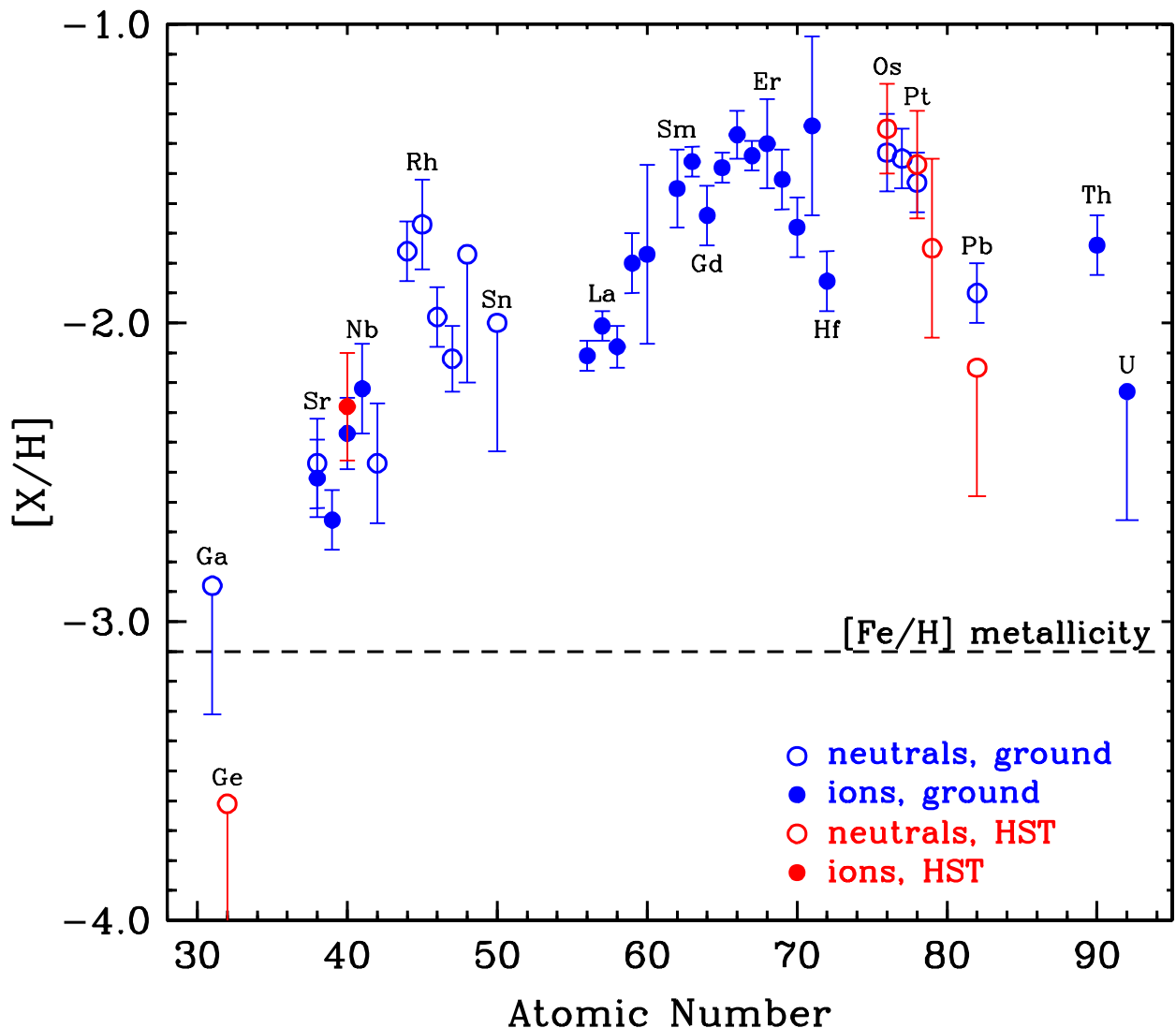}
\caption{
Abundances with respect to the Sun [X/H] versus atomic number for elements
with Z~$>$~30.
The symbols are defined in the figure legend, and as in Figure~\ref{f1}
a dashed line represents the [Fe/H] metallicity of \cs22.
Detected elements are shown with complete error bars, and upper limits are 
shown with only the lower half of an arbitrary-length error bar.
A few element symbols are marked to help with identification of the points.
\label{f3}}
\end{figure}

The derived abundances of the $n$-capture elements are listed in
Table~5, and their [X/H] values are displayed in
Figure~\ref{f3}.
The relation between measured $\sigma$ values and adopted error bars
in this table and remaining figures for these elements follow the rules 
described previously for the lighter elements.
The general shape of this abundance distribution, increasing [X/H] 
with increasing atomic number up to Z~$\approx$~65, is not new and
can be compared to a similar plot in Figure~4 of Sneden
\etal\ (1996)\nocite{sne96}.
But the present results are based on 36 abundances or significant upper
limits compared with the 20 reported by Sneden et~al. 
The total number of transitions involved is 299 compared to the earlier 125. 
For species with two or more transitions, the median of the sample 
deviations $\sigma$ is 0.10 (Table~5) compared with 
0.15 (Table~2 of Sneden \etal).

When an abundance has been determined from many lines (N~$\gtrsim$~5)
its internal abundance uncertainty is small.
Moreover, intercomparisons of any two $n$-capture elemental abundances 
derived from the same species (\eg, [\ion{Ru}{1}/\ion{Pd}{1}]
or [\ion{La}{2}/\ion{Eu}{2}]) are likely to have small external errors
because atmospheric parameter uncertainties produce (roughly) the same
effects on both elements in the ratios.
We have argued in past papers (\eg, Sneden \etal\ 1996\nocite{sne96})
that LTE analyses should be reasonable approximations for the
$n$-capture elemental abundances derived from ionized transitions.
Additionally, to the extent that possible departures from LTE affect
ionization equilibria in similar ways, the general agreement between 
abundances from neutral and ionized transitions of six Fe-peak elements 
(Table~4) and of Sr (Table~5) lends confidence 
in the general $n$-capture abundance results derived from neutral species.
Statistical equilibrium computations could potentially settle such questions,
but are beyond the scope of this paper.
Indeed these ``NLTE'' studies probably are not possible for each element
at present, due to the incompleteness of atomic data for many species.

Our earlier papers on \cs22\ have included discussions of elements with
multiple detected transitions for which abundance computations are relatively
straightforward.  
As one example, consider La. 
Parameters of \ion{La}{2} lines have been taken from a recent comprehensive 
lab study (Lawler \etal\ 2001\nocite{law01}). 
For \cs22, there are 15 lines contributing to the mean La abundance, with
very small line-to-line scatter ($\sigma$~= 0.05, Table~5).
We regard the La abundance as secure, and little comment is needed 
for this and many other elements.

In the next two subsections we will focus attention on more difficult 
situations: abundances and upper limits determined from one or two spectral 
features.
These abundances cannot be employed with the same confidence
as can those of ``the usual suspects'' among the $n$-capture elements.  
The cases of Mo and Sn will be used as illustrative examples to show
the prospects and perils of the new element search we have conducted
for \cs22.
In a third subsection comments will be given as needed for other $n$-capture 
element abundance analyses.

\subsubsection{Detection of Molybdenum}

In the visible and near-UV spectral regions only \ion{Mo}{1} lines
will be detectable in cool, metal-poor stars; strong \ion{Mo}{2}
lines occur only in the vacuum UV.
We searched the transition probability compendium of Whaling \& Brault 
(1988)\nocite{wha88} for promising \ion{Mo}{1} lines, and found 
about 10 of them that merited closer investigation.
Frustratingly, all but one of these lines proved to be either undetectably
weak or severely blended in our \cs22\ spectra.  
The unusable transitions include all the ones that were employed in
the Bi\'emont \etal\ (1983)\nocite{bie83} solar Mo abundance study.
The elimination process left only the 3864.10~\AA\ line.

There is an absorption clearly visible at 3864.10~\AA\ in the \cs22\
spectrum, but is it due to \ion{Mo}{1}?
The Mo contribution to this feature is very difficult to detect in most cool 
(\teff~$<$ 6000~K) stars, because it is intrinsically weak and is buried 
beneath strong CN $B^{\rm 2}\Sigma^+-X^{\rm 2}\Sigma^+$ violet 
system (0-0) and (1-1) band lines.
Only in the relatively rare stars like \cs22\ can low metallicity and
large departures from the Solar abundance mix potentially combine to 
reveal the presence of the \ion{Mo}{1} contribution.
But there are also other potential contaminants in this very crowded
spectral region, including lines of the CH $B^{\rm 2}\Pi-X^{\rm 2}\Pi$
system as well as Fe-peak and $n$-capture atomic features.
Disentangling all these potential contributors to the 3864.10~\AA\ 
feature is another process of elimination.

To help in this effort, we compared our \cs22\ spectrum to that of \bd17
(Cowan \etal\ 2002\nocite{cow02}), an $r$-process-rich star with 
somewhat different characteristics than \cs22.  
Forming differences $\delta$ in the sense 
\bd17\ $minus$ \cs22, first note that \bd17\ is hotter 
($\delta$\teff~= +400~K) and so will have intrinsically weaker spectral 
absorption lines for any element with equal abundances in the two stars.
Second, \bd17\ is less metal-poor ($\delta$[Fe/H]~= +1.0).
Third, the $n$-capture abundance patterns are essentially identical in these 
two stars, but $\delta$[Eu/H]~= +0.3 (the larger metallicity of \bd17\ wins 
out over the larger relative $n$-capture enhancements of \cs22).
Fourth, C is more abundant in \cs22 ($\delta$[C/H]~= --0.4), but N is slightly 
more abundant in \bd17\ ($\delta$[N/H]~= +0.2).
These differences in effective temperature and abundances lead to substantial
differences in the spectra near the \ion{Mo}{1} 3864.10~\AA\ line.

\begin{figure}
\epsscale{0.9}
\plotone{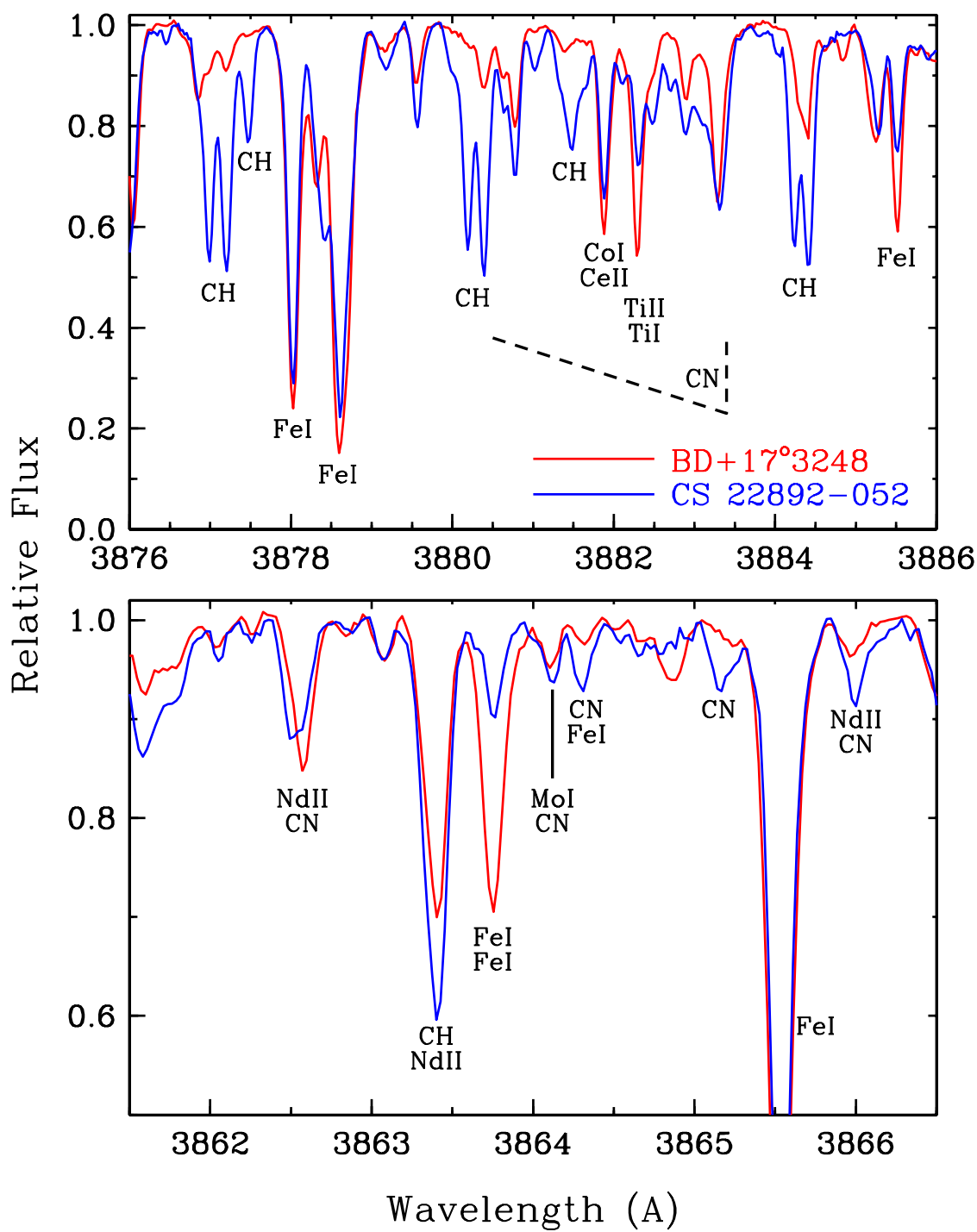}
\caption{
Spectra of \cs22\ and \bd17\ (Cowan \etal\ 2002)
in two small spectral regions relevant to detection of the
\ion{Mo}{1} 3864.10~\AA\ line.
In the top panel the wavelength region near the
CN $B^{\rm 2}\Sigma^+-X^{\rm 2}\Sigma^+$ (0-0) bandhead is shown,
and in the bottom panel a small portion of spectrum surrounding the
\ion{Mo}{1} line is shown.
A few major feature identifications are labeled in each panel, and
the growth of CN strength with increasing wavelength toward the
3883.4~\AA\ bandhead is indicated by dashed lines in the top panel.
\label{f4}}
\end{figure}

In Figure~\ref{f4} two small spectral regions of \cs22\ and \bd17\ are 
displayed.
In the top panel, a 10~\AA\ segment surrounding the CN violet system 
(0-0) bandhead illustrates the differences of molecular and Fe-peak line 
absorptions in the two stars.
The CH lines of \cs22\ are 5-10 times stronger than in they are in \bd17,
a combined consequence of its larger C abundance and lower \teff.
The ratio of CN strengths is about a factor of three at the bandhead (the
\teff\ difference is the cause here, as the C+N abundances are similar 
in the two stars). 
Away from the bandhead the CN lines are very weak in both stars.
The Fe-peak lines are 1.5-2 times stronger in \bd17, due to its higher 
metallicity.

As shown in the bottom panel of Figure~\ref{f4}, the absorption in the 
two stars at the \ion{Mo}{1} wavelength is nearly identical.
This equality suggests which species probably are $not$ dominating 
the feature: CH and atomic lines of elements with Z~$\leq$~30.
If the 3864.10~\AA\ feature is caused by CH, the absorption in \cs22\ 
would be overwhelmingly larger than in \bd17. 
If it is caused by a line of a ``light'' element, the \bd17\ spectrum 
should have the stronger absorption.
A conspiracy between CH and atomic contaminants creating similar 3864.10~\AA\
lines in the two stars cannot be totally ruled out, but our synthetic
spectrum line list culled from the Kurucz (1995)\nocite{kur95}
database has no neutral and ionized lines of Z~$\leq$~30 elements
within $\pm$0.09~\AA\ of this wavelength.

As indicated in the bottom panel of Figure~\ref{f4}, some CN 
absorption should exist in the 3860-3868~\AA\ region of the \cs22\ 
spectrum. 
The positive identification of CN features from our synthetic
spectrum calculations is supported by the much weaker absorption at these
wavelengths in the \bd17\ spectrum.
Trial synthetic spectra that forced agreement with the observed features
at 3864.3, 3865.2, and 3866.0~\AA\ suggest three related points: 
$(a)$ the implied N abundance is $\sim$0.5~dex larger than that
given in Table~4, which was derived from the 3883~\AA\
bandhead; $(b)$ this same N abundance produces absorptions that are 
obviously too strong for the observed CN features in neighboring spectral
regions; and $(c)$ the CN part of the 3864.10~\AA\ feature
is at most 20-30\% of the total absorption.

If the absorption at 3864.1~\AA\ is attributable to neither CH nor lighter 
atomic species, and has only a small CN contribution, it could
be due to an $n$-capture species other than \ion{Mo}{1}.
Two possible such lines are known.
First, Palmeri \etal\ (2000)\nocite{pal00} list a \ion{Ce}{2} line
at 3864.107~\AA, but the combination of its EP~= 0.96~eV and
log~$gf$~= --2.04 render its absorption undetectably small unless the
Ce abundance is a factor of $\sim$10 larger than we have derived
(Table~5).
Second, Kurucz (1995)\nocite{kur95} lists a \ion{Sm}{2} line at
3864.047~\AA, with EP~= 0.66~eV and log~$gf$~= -1.15.
These line parameters are more favorable, and increasing our derived
Sm abundance by only 2-3 times produced detectable absorption.
But this distorts the line profile to the blue by unacceptable amounts.
Thus the other known $n$-capture contributors to the 3864.1~\AA\ feature
are insignificant contaminants to the \ion{Mo}{1} line.

With attention to the above issues, we have derived 
log~$\epsilon$(Mo)~= --0.55.
Tests with adding or subtracting contamination from other species suggests
that the 3864.1~\AA\ feature can be matched by Mo abundances that vary
by at most 0.2~dex, yielding in particular an upper limit to the abundance
of $\approx$--0.3.
We attach the estimated 0.2~dex uncertainty to the Mo value in all 
abundance plots to follow.

\subsubsection{Non-detection of Tin}

A search was made for lines of \ion{Sn}{1} in the manner described above
for \ion{Mo}{1}; strong lines of \ion{Sn}{2} occur only in the vacuum $UV$.  
In the visible and near-$UV$ spectral regions, the only two that appeared
to be potentially detectable and relatively unblended lie at 3262.34 and 
3801.02~\AA.
Atomic and molecular line lists for several \AA\ surrounding these two
lines were assembled in the usual manner, and applied to our observed
(Keck) spectra of \cs22.
Comparisons of synthetic and observed spectra are displayed in
Figure~\ref{f6}.
Neither \ion{Sn}{1} line is clearly present in our data.

These two lines do not present ideal cases for Sn detection, as each
is clearly contaminated by other absorptions.
The blending agents for the 3262~\AA\ line are \ion{Sm}{2} 3262.27~\AA\ 
and \ion{Os}{1} 3262.29~\AA; their contributions to the observed feature
in the upper panel of Figure~\ref{f6} are roughly equal.
Repeated trial synthetic spectra demonstrated that the \ion{Sn}{1} line
cannot account for this feature, mainly because it lies 0.05~\AA\ to the 
red of the observed absorption.
But our synthetic spectrum in this small wavelength interval does not
adequately reproduce several other observed features, which does not
inspire confidence in a detailed analysis of the 3262.33~\AA\ line. 
A better overall match between observed and synthetic spectra is
seen in the bottom panel of Figure~\ref{f6} for the 3801.02~\AA\ line.
The \ion{Sn}{1} line is bounded to the blue mainly by \ion{Sm}{2} 
3800.89~\AA\ and to the red by \ion{Nd}{2} 3801.12~\AA.
Their absorptions are clearly present but that of \ion{Sn}{1} is not 
obvious in the observed \cs22\ spectrum.

Our estimation that log~$\epsilon$(Sn)~$<$ 0.0 is strengthened from
the non-detection of $both$ \ion{Sn}{1} lines. 
If only one line had been available for analysis the abundance limit
would probably have been higher, with larger uncertainty.

\subsubsection{Other $n$-Capture Abundances}

$Sr, Ba, and~Yb:$  The atomic structures of the ionized
species of these elements produce 2-4 extremely strong resonance and
low-excitation transitions (\eg, \ion{Sr}{2} 4077.7~\AA, \ion{Ba}{2}
6141.7~\AA).
Analyses of these features dominate the large abundance literature for Ba 
and Sr. 
The occurrence of both \ion{Yb}{2} resonance lines below 3700~\AA\ 
has severely limited studies of this element.
All of these lines lie on the flat or damping part of the curve-of-growth
in the spectra of $n$-capture-rich metal-poor giants.
These elements have multiple isotopes, of which the odd-mass ones have
substantial hfs.
Both of these effects must be carefully accounted for in synthetic
spectrum computations, and still the derived abundances from such deep 
and saturated absorption features are dependent on assumed values of 
microturbulent velocity.
In the present analysis, however, we have been able to detect two
very weak high-excitation lines of \ion{Sr}{2} and the weak resonance
line of \ion{Sr}{1} (Table~3).
These transitions yield abundances that are consistent with the two very
strong \ion{Sr}{2} resonance lines.
Abundances from three high-excitation \ion{Ba}{2} lines agree with
the results from the low-excitation features.\footnote{
The \ion{Ba}{2} 4934.10~\AA\ line is rarely employed in Ba abundance
studies because it is blended (mainly) with a strong \ion{Fe}{1} line at
4934.08~\AA.
This contaminant was included in our synthetic spectrum computations,
but proved to be a negligible contributor in \cs22.
}
The abundances from the strong low-excitation ionized lines appear
to be reliable in \cs22.

$Cd:$ Sneden \etal\ (2000)\nocite{sne00a} claimed detection of
this element, and they derived log~$\epsilon$(Cd)~= --0.35~$\pm$~0.20 from
analysis of just the \ion{Cd}{1} 3261.05~\AA\ transition.
We re-examined the Cd abundance, and confirmed that there are no other
\ion{Cd}{1} or \ion{Cd}{2} lines strong enough to be useful abundance
indicators in \cs22.
However, our new exploration of the 3261.05~\AA\ line highlighted the 
possibly substantial contribution of an OH line at the same wavelength 
to the total absorption feature.  
If the OH line is neglected we derive log~$\epsilon$(Cd)~$\sim$ 0.0.
Inclusion of the OH at a strength consistent with other probable
OH lines in this spectral region suggests that it might dominate the 
3261~\AA\ contribution. 
Therefore we have chosen in Table~5 and in $n$-capture
abundance figures to enter log~$\epsilon$(Cd)~$<$ 0.0.
This abundance limit should be viewed with caution.

$Ge, Y, Os, Pt, Au, and~Pb:$  Abundances for these elements are wholly
or partly determined from our HST spectrum.
The general concordance of HST-based abundances of Y, Os, and Pt with
those determined from ground-based data increases confidence in
the HST data, even though it has relatively poor S/N.
The very low abundance of Ge determined from the HST spectrum
of the \ion{Ge}{1} 3039.07~\AA\ line gains some support from the
the low Ga abundance implied from the non-detection of the 
\ion{Ga}{1} 4172.04~\AA\ feature in our Keck and VLT spectra.
Finally, we are not confident of the suggested detections of two \ion{Pb}{1} 
lines in the ground-based spectra. 
These lines should be nearly ten times weaker than the 2833.05~\AA\ line,
which we cannot detect in the HST spectrum. 
The derived Pb abundance upper limit from the 2833~\AA\ line may well be more
reliable than the abundances determined from the questionable detections
of the other two \ion{Pb}{1} lines.

\section{DISCUSSION}

\subsection{Neutron-Capture Abundance Comparisons}

Our new observations have provided abundances or upper limits for 36
{\it n}-capture elements, including seven more (Ga, Ge, Mo, Sn, Lu, Pt, Au) 
not seen in our previous study of \cs22\ (Sneden \etal\ 2000a). 
This provides more abundances than any other halo star, or in fact any 
other star except the Sun.

\subsubsection {Heavier N-Capture Elements}

In  Figure~\ref{f6} we compare these abundances in \cs22 with the 
(scaled) solar system $r$-process abundance curve.
This distribution, indicated by the solid line, is based upon 
$n$-capture cross section measurements 
(K\"appeler \etal\ 1989\nocite{kap89}; Wisshak \etal\ 1996\nocite{wis96})
and assumes the ``classical'' $s$-process empirical relation between
abundance and cross section. 
The resulting $r$-process isotopic abundances are the difference between the 
solar value and the individual $s$-process isotopic abundances. 
Summing the isotopic abundances results in the $r$-process elemental 
abundance distribution (Burris \etal\ 2000\nocite{bur00}).
It is clear from Figure~\ref{f6} that the relative abundances of 
the heaviest $n$-capture elements, 56~$\leq$~Z~$\leq$~82, are all consistent 
with the solar system $r$-process curve. 
This confirms and extends the previous results in \cs22\ (Sneden \etal\ 
2000\nocite{sne00a}) and in other stars (Westin \etal\ 2000\nocite{wes00};
Johnson \& Bolte 2001\nocite{joh01}; Cowan \etal\ 2002\nocite{cow02}; 
Hill \etal\ 2002\nocite{hil02}). 
However, for \cs22 and the majority of the other cases there was very 
limited data of the heaviest \third\ $r$-process peak elements, with 
dominant transitions in the $UV$ necessitating HST observations.  
The agreement between the $n$-capture abundances in this old halo star
and the solar distribution strongly supports the claim of a robust $r$-process 
over the history of the Galaxy, at least for the elements Ba and above.

Support for this conclusion is also found in recent isotopic abundance studies. 
Sneden \etal\ (2002) found that the \iso{151}{Eu} and \iso{153}{Eu}
isotopes in three halo stars, including \cs22, are in agreement with each 
other and in solar proportions.
From an analysis of another metal-poor star (HD 140283), Lambert 
\& Allende-Prieto (2002)\nocite{lam02} find that the isotopic fractions of 
Ba are also consistent with the solar $r$-process values.

As one further indication of the abundance history of these elements prior
to the formation of \cs22, we also show in Figure~\ref{f6} an abundance 
comparison with the solar system $s$-process elemental distribution 
(dashed line) from Burris \etal\ (2000). 
This comparison demonstrates clearly that these elements observed in 
\cs22, including ``traditional'' $s$-process elements such as Ba,
were synthesized solely in the $r$-process and not in the $s$-process in a 
previous stellar generation early in the history of the Galaxy. 

A detailed element-by-element difference comparison between the solar
$r$-process elemental curve from Burris \etal\ (2000) and the abundances 
in \cs22\ is presented in the top panel of Figure~\ref{f7}. 
The mean difference between the solar curve and the \cs22\ abundances (in 
log~$\epsilon$ units) for 56~$\leq$~Z~$\leq$~79 is --1.42~$\pm$~0.02
($\sigma$~= 0.10; see Table~6).
That is, the mean stellar abundance level of these elements is deficient 
with respect to the solar $r$-process abundance set by a factor of 
$\simeq$~25, but enhanced with respect to the \cs22\ Fe abundance  
by a factor of $\simeq$~45.
The very small $\sigma$ is consistent with stellar and solar abundance
uncertainties.
In addition to the curve from Burris \etal, there have been
other solar {\it r}-process abundance determinations. 
Rather than assuming the model independent, but empirical, ``classical'' 
$s$-process relationship, more sophisticated stellar models based upon
$s$-process nucleosynthesis in low-mass AGB stars have been developed
to predict the isotopic $s$-process contributions to the solar abundances 
(Arlandini \etal\ 1999\nocite{arl99}).  
In the lower panel of Figure~\ref{f7} we make a similar comparison 
between the Arlandini \etal\ solar system $r$-process curve
and the abundances in \cs22. 
The mean difference and its uncertainty are nearly identical to
the comparison with the Burris \etal\ data.
This testifies to the success of the Arlandini \etal\ model in fitting 
the solar system heavy element $s$-process abundances. 

While the overall agreement between the abundances in \cs22\ and these two
solar system $r$-process curves is excellent, there are some small 
deviations for a few elements (\eg, Lu and Au). 
It might be important to examine additional cases with such deviations.   
In some earlier studies small differences between the 
elemental abundances and the solar curve were the result of  
inadequate atomic physics data (\eg, oscillator strengths). 
Most of these anomalies have been corrected by recent lab atomic
data studies.
We also note that for cases for which the $s$-process contributions to a
solar elemental abundance completely dominate the $r$-process contributions, 
the uncertainties associated with the extracted $r$-process residuals can 
be extremely large.

\subsubsection {Lighter $n$-Capture Elements}

It is clear from examination of Figure~\ref{f6} and
Figure~\ref{f7} that some of the lighter {\it n}-capture elements, those 
below Ba, show significant deviations from the solar $r$-process curve.  
Until recently there have been little data available on the lighter
$n$-capture elements, particularly those in the range 41~$\leq$~Z~$\leq$~55.
Abundance results for \bd17\ (Cowan \etal\ 2002\nocite{cow02}) indicate 
that a few of the elements in this regime, specifically the element Ag, 
appear to deviate from the same solar curve that fits the heavier 
$n$-capture elements.
This perhaps supports an earlier suggestion, based upon solar system
meteoritic studies, of two $r$-processes --- one for the  elements
A $\gtrsim$ 130-140 and a second $r$-process for the lighter elements
(Wasserburg, Busso \& Gallino 1996\nocite{was96}).
As illustrated in Figure~\ref{f6} and Figure~\ref{f7},  
we have now detected 5 elements in this lighter element regime for \cs22,
and also obtained two significant abundance upper limits.
Note that Ag and Mo fall well below the scaled solar $r$-process curve. 
The abundance of Pd also falls below the solar curve, while both Nb and Rh 
lie on, or very near this distribution. 
The upper limit on Cd (and Sn with respect to the Arlandini 
\etal\ predictions) could also be consistent with the heavier 
$n$-capture abundances. 
On average, however, these lighter elements do seem to have been 
synthesized at a lower abundance level than the heavier $n$-capture elements. 
The mean difference between the solar curve and the \cs22\ abundances 
for 38~$\leq$~Z~$\leq$~55 is --1.70 and --1.65 with a scatter 
$\sigma$ about the mean of 0.26 and 0.22 with respect to Burris 
\etal\ (2000) and Arlandini \etal, respectively (Table~6).

There are several possible explanations for the differences in 
the abundance data for the lighter and heavier {\it n}-capture elements.
It has been suggested for example that perhaps, analogously to the 
$s$-process, the lighter elements might be synthesized in a ``weak'' 
{\it r}-process with the heavier elements synthesized in a  more robust 
``strong'' (or ``main'') {\it r}-process.   
Supernovae of a different mass range or frequency (Wasserburg \& Qian 2000)
or the helium zone of an exploding supernovae (Truran \& Cowan 2000),
have been suggested as possible second {\it r}-process sites that might 
be responsible for the synthesis of nuclei with A $\lesssim$ 130--140.
Alternative interpretations have suggested that that the entire abundance 
distribution could be synthesized in a single core-collapse supernova 
(Sneden \etal\ 2000a; Cameron 2001).

We also note that the upper limits on Ga and Ge are significantly below
the solar {\it r}-process curve. 
These elements have significant contributions from $n$-capture 
synthesis (Burris \etal\ 2000).
However, the very low upper limits are more suggestive of an abundance 
that scales with the very low iron abundance in this star, similar to 
that seen in several other metal-poor halo stars (Sneden \etal\ 1998). 
Additional studies of the nature of the synthesis of Ga and Ge 
in other stars are currently underway (Cowan \etal\ 2003).

\subsection {Theoretical Predictions}

Our reported observations of the heavy element abundance patterns in 
\cs22\ reveal and confirm two important features that are common 
to extremely metal deficient stars: (1) the robustness of the goodness 
of fit to solar system $r$-process abundances, for elements in the range 
from barium to lead; and (2) the underproduction of $r$-process elements 
in the region of mass number below barium (A~$\lesssim$ 130-140) relative 
to the barium-to-lead element region. 
We proceed now on the assumption that the robustness in the heavy 
region continues through the actinides, so that we can utilize abundance 
data concerning the interesting actinide radioactivities $^{232}$Th, 
$^{235}$U, and $^{238}$U to date the star.
A potential test of this robustness follows from theoretical calculations of 
$r$-process synthesis. 
Specifically, a better understanding of both of the features identified 
above can be gained by comparison of representative calculations of 
$r$-process nucleosynthesis with the abundances determined for \cs22. Note again that the large enhancement of $r$-process products 
relative to iron in \cs22\ makes this a particularly appropriate choice 
for this kind of analysis.

Abundance comparisons with theory are shown in Figure~\ref{f8}, 
for two specific $r$-process calculations. 
Both are based upon the equilibrium (i.e., ``waiting 
point approximation'') model for the $r$-process. The first, shown in the top
panel of Figure~\ref{f8}, was presented and discussed by 
Cowan \etal\ (1999)\nocite{cow99}. 
These calculations utilized a mass formula based upon the ETFSI-Q 
(Extended Thomas Fermi with Strutinski Integral and Quenching) model. 
The calculations consisted of a 16 component least-squares fit to the 
solar abundance distribution in the mass range 80~$\leq$~A~$\leq$~208. 
For the purpose of this paper, these calculations have now been updated 
and improved, with the inclusion of recent experimental determinations of 
half lives and neutron pairing energies (Pfeiffer, Kratz \&  Moeller 
2002\nocite{pfe02}; Moeller, Pfeiffer \& Kratz 2002\nocite{moe02}). 
The calculations have then been fit to the solar $r$-process abundances in
the mass range 125~$\leq$~A~$\leq$~209.
Comparisons of these new theoretical results with \cs22\ abundances are 
shown in the lower panel of Figure~\ref{f8}. 

It is clear from the comparisons summarized in both panels of 
Figure~\ref{f8} that the agreement with $r$-process in the 
mass region (130~$\lesssim$ A~$\lesssim$ 208) is extremely good, with 
a mean offset of -1.3, similar to that found for Burris \etal\ and 
Arlandini \etal\ (see Table~6).
The scatter about the mean for the theoretical predictions is somewhat
larger, when compared to the other solar comparisons, with $\sigma$~=~0.26.  
The theoretical comparisons again confirm a significantly lower
abundance level for the lighter {\it n}-capture elements
in \cs22\ with respect to solar {\it r}-process curve --
the mean differences and $\sigma$ values very close  to what was found 
in the Burris \etal\ and Arlandini \etal\ comparisons. 

The excellent agreement between theory and observed heavy element $r$-process 
abundances gives us some confidence in proceeding to utilize the Th and U 
chronometers to provide an age estimate for \cs22.

\subsection {Cosmochronometry}

We now employ the radioactive abundances to make
cosmochronometric age estimates for this star. 
There have been a number of recent detections of Th in halo and globular
cluster stars (Sneden \etal\ 1996, 2000b; Cowan \etal\ 1999, 2002; 
Westin \etal\ 2000; Johnson \& Bolte 2001; Cayrel \etal\ 2001).
Comparison of the observed stellar abundances of the radioactive elements 
and the initial abundances of these elements during $r$-process synthesis
leads to a direct radioactive age determination.
These chronometric age estimates, however, depend sensitively upon the 
predicted initial values of the radioactive elements, in ratio to each 
other, or to stable elements. 
To determine these initial ratio values we have utilized the theoretical
$r$-process predictions described above in \S 4.2.
Constrained by reproducing the observed stable element abundances observed in 
\cs22, the earlier calculations (Cowan \etal\ 1999) suggested an initial 
value, resulting from $r$-process nucleosynthesis, of Th/Eu = 0.48.
Comparing the observed Th/Eu ratio to this theoretically predicted 
(time-zero) Th/Eu value yields a chronometric age of 14.1 Gyr.
The newer refined theoretical calculations, including more experimental 
values, yields an initial value of Th/Eu = 0.42, and leads to a chronometric 
age estimate of 11.4~Gyr. 
This spread in ages demonstrates the sensitivity of the 
radioactive dating method and provides a rough estimate of 
the uncertainties associated with the theoretical predictions from 
some of the best available $r$-process mass formulae and models. 
As illustrated in Figure~\ref{f7}, the Eu abundance value lies 
slightly below (by 0.04 in log~$\epsilon$) the mean fit to the 
rare earth elements -- this is a measure of the observational uncertainty, 
and in this case would lead to an increase in the age estimates of 
approximately 1.8~Gyr. 

We show in Table~7 the mean age estimates for \cs22, where 
we have taken the average predicted values from the two model calculations 
in comparison to the newly-determined oberved abundance values. 
Thus, for example, this averaging yields an age of 12.8 Gyr based upon 
the Th/Eu chronometer.   
We can also employ the abundances of some of the third $r$-process-peak 
elements for age estimates. 
Ignoring the relatively uncertain Os abundance for these 
estimates, averages for the chronometer pairs Th/Ir and Th/Pt 
suggest ages ranging from 10.5 to 19.2 Gyr (see Table~7) . 
An average of those values suggests an overall chronometric estimate of
14.2~$\pm$~$\simeq$~3~Gyr.   
These chronometric values can be compared with our previous age estimate 
of 15.6~$\pm$~4~Gyr (Sneden \etal\ 2000a) for this star -- the differences 
resulting from slightly different abundance determinations and different 
theoretical predictions for the initial abundances. 

As an additional indication of the age of this star we list in 
Table~7 the solar system ratios for the various chronometers. 
These values are measured and do not depend upon theoretical predictions 
and their associated nuclear physics uncertainties. 
Thus, we can compare the solar values (as an indication of the initial zero 
decay-age $r$-process abundance ratios) with the observed stellar values.
However, since the Sun is only approximately 4.5 Gyr old, the solar ratios 
do include Galactic chemical evolution effects and represent a lower limit 
to the zero decay-age $r$-process abundances.
This follows since Eu is stable but Th (while constantly produced and 
ejected into the interstellar medium) has partially decayed in the time
between the formation of the oldest stars and the Sun.
This leads, in general, to lower (with respect to theoretical zero decay-age 
$r$-process) values of Th/(stable elements) for solar ratios.
Therefore, the ages derived with the solar ratios should be viewed formally
as lower limits.
We note, however, that there are cases where some of the predicted ages 
are less than the expected lower limit solar system age estimates. 
This is an indication that some of the theoretically predicted abundances 
need to be revised -- for example, an overestimate of the abundance of 
Pt and corresponding underestimate for Ir.  
Such revisions are currently underway (Burles \etal\ 2003).
Nevertheless, there is a large overlap in the lower limit age estimates 
with those based upon theoretical $r$-process predictions.  
An average of the chronometer pairs, assuming initial solar system ratios,  
gives an age of 14.7~Gyr, which is not inconsistent with the average 
based upon theoretically predicted $r$-process abundance ratios.

While Th/Eu has been utilized extensively  for chronometric age estimates 
(Sneden \etal\ 1996,2000; Cowan \etal\ 1999, 2002; Johnson \& Bolte
2001) this ratio gives a very unrealistic age estimate in CS 31082--001 
(Cayrel \etal\ 2001; Hill \etal\ 2002; Schatz \etal\ 2002\nocite{sch02}; 
Wanajo \etal\ 2002\nocite{wan02}).
The detection of U in that star allows for a second $r$-process chronometer 
and suggests an age of 14.1~$\pm$~2.5~Gyr (Wanajo \etal) to 
15.5~$\pm$~3.2~Gyr (Schatz \etal).  
Since Th/Eu and Th/U give similar age results (13.8 $\pm$ 4 Gyr) 
for \bd17\ (Cowan \etal\ 2002), 
it is not clear yet why CS~31082--001 is so different.
It does suggest that Th/U is a more reliable chronometer --- particularly
since they are nearer together in mass number than, for example, Th and Eu. 
Unfortunately we are only able to determine an upper limit on U in \cs22. 
We can employ the chronometer ratio of Th/(upper limit of U) to determine 
a lower limit on the age of \cs22.
Recent calculations suggest a possible range of the initial Th/U value of 
1.45 to 1.81 (Burles \etal\ 2003), leading to limiting ages for this star  
of 12.4 and 10.4 Gyr, respectively.  
This implies a conservative lower limit of 10.4 Gyr, based solely upon the 
abundance of Th and the upper limit  on the U abundance.  
While these values are not definitive, they are consistent with the 
other age determinations employing, for example, Th/Eu. 
It would be extremely important for chronometric studies to eventually 
obtain a reliable U abundance in this star.

\section{CONCLUSIONS}

We have conducted an extensive new abundance analysis of the
very-metal-poor, $n$-capture-rich, halo giant star \cs22.
We have employed high resolution spectra obtained with echelle
spectrographs of HST, Keck~I, McDonald, and VLT.
We have obtained abundances for 52 elements and meaningful upper
limits for five more elements, the largest set of elements
analyzed in any star other than the Sun.
New detections in \cs22\ include the $n$-capture elements Mo, Lu, Au,
Pt and Pb, and upper limits encompass Ga, Ge, Cd, Sn, and U.

We have derived abundances for both ionization stages of six Fe-peak
elements, lending confidence to the derived model atmosphere for
\cs22.
Abundances of the volatile light elements Li and Be indicate large
depletion factors ($>$10) for both elements, confirming the status
of this star as an evolved, convectively mixed giant.

Close matches between observed abundances of stable heavy (Z~$>$~56)
elements and two empirical scaled solar $r$-process abundance distributions
re-enforces the suggested universality of the $r$-process in this
element regime.
In general, less satisfactory matches are seen for lighter $n$-capture
elements -- this may indicate the existence of two $r$-process sites
or multiple $r$-process production parameters in single sites.
Scaled theoretical $r$-process predictions give a reasonable fit to
\cs22\ abundances (albeit with large element-to-element scatter), 
providing some encouragement that the predicted production ratios for 
Th/U and Th/Eu are realistic. 

For Ga and Ge, two elements at the boundary between charged-particle
fusion and $n$-capture synthesis modes, we derive upper limits much more
consistent with the (very low) Fe-peak abundances than with the
(enhanced) $n$-capture abundances.
It would thus appear that, at least for this and some other old halo 
stars, $r$-process synthesis has not contributed significantly to 
these elements. We note that the $r$-process contributions to these lighter 
elements may have their origin, for example, in a quite different  
$r$-process site, in stars of longer lifetimes than those whose abundance
products enriched the early Galactic halo gas.

The abundance of Th and the U upper limit, and the abundance of 
Th with respect to those of stable $n$-capture elements in \cs22, 
yield self-consistent chronometric ages. This is encouraging particularly 
in light of the fact that this appears not to be true for the halo 
star CS~31082-001 (Hill \etal \ 2002). 
The detection of Th and the upper limit on the U abundance together imply
a lower limit of 10.4 Gyr on the age of \cs22, quite consistent with
the Th/Eu age estimate of 12.8 $\pm$ $\simeq$ 3 Gyr.
An average of several chronometric ratios 
suggests an overall chronometric age estimate of
14.2 $\pm$ $\simeq$ 3 Gyr for \cs22. 

Our study comes close to exhausting the element detection possibilities
for \cs22.
Atomic physics limitations suggest that one is not likely to detect 
many additional elements for this star in the near future.
However, higher resolution and S/N spectra may be able to turn some
of the upper limits reported here into detections.
Renewed attempts to detect U would be particularly useful.
Very detailed abundance distributions for additional $n$-capture-rich stars
will be needed to explore further the questions raised by the present study.

\acknowledgments

We thank Roger Cayrel, Con Deliyannis, Peter H\"oflich, David Lambert, 
Caty Pilachowski, Craig Wheeler, and the referee for helpful discussions 
and suggestions for improvement of this paper.
This research has been supported in part by STScI grant GO-08342, NSF grants 
AST-9986974 (JJC), AST-9987162 (CS), AST-0098508 and AST-0098549 (TCB),
AST-9819400 (JEL), PHY-9800980 (GMF), and at the University of Chicago (JWT)
by DOE contract B341495 to the ASCI/Alliances Center for Astrophysical  
Thermonuclear Flashes and DOE contract DE-FG02-91ER40606 in Nuclear Physics 
and Astrophysics.  
Support for BP and KLK was provided by the German  BMBF (grant 06MZ9631).
Support for BP and KLK was provided by the German  BMBF (grant 06MZ9631).
Research for III is currently supported by NASA through Hubble Fellowship
grant HST-HF-01151.01-A from the Space Telescope Science Institute, which
is operated by the Association of Universities for Research in Astronomy,
Incorporated, under NASA contract NAS5-26555.
III is also pleased to acknowledge earlier financial support for this work
from the University of Texas, through McDonald Observatory and through
Graduate Fellowships.
This research has made use of the NASA/IPAC Infrared Science
Archive, which is operated by the Jet Propulsion Laboratory,
California Institute of Technology, under contract with NASA and NSF.
JJC thanks the University of Texas at Austin Department of
Astronomy John W. Cox Fund for partial support while this
paper was being written.

\newpage

\appendix

\section{NEW ELEMENTS IN THE \cs22\ SPECTRUM}

Many laboratory studies have contributed to the $n$-capture line lists
employed in the \cs22\ abundance analysis.
We refer the the reader to our previous papers for discussions of
the following species:
{\it (a)} \ion{Ge}{1}, \ion{Nb}{2}, \ion{Pd}{1}, \ion{La}{2}, \ion{Ce}{2}, 
          \ion{Pr}{2}, \ion{Nd}{2}, \ion{Eu}{2}, \ion{Gd}{2}, \ion{Tb}{2}, 
          \ion{Dy}{2}, \ion{Os}{1}, \ion{Ir}{1}, \ion{Pt}{1}, \ion{Au}{1}, 
          and \ion{U}{2}: Cowan \etal\ (2002)\nocite{cow02}.
{\it (b)} \ion{Sr}{1}, \ion{Sr}{2}, \ion{Y}{2}, \ion{Zr}{2}, \ion{Ba}{2}, 
          \ion{Sm}{2}, \ion{Er}{2}, and \ion{Tm}{2}:
          Sneden \etal\ (1996)\nocite{sne96}.
{\it (c)} \ion{Os}{1}, \ion{Pt}{1}, and \ion{Pb}{1}:
          Sneden \etal\ (1998)\nocite{sne98}.
{\it (d)} \ion{Ag}{1}: Crawford \etal\ (1996)\nocite{cra98}.
Here we discuss newly detected elements and revised data for previously
known transitions in \cs22.

{\em \ion{Ga}{1}:} The $gf$ value of the 4172.04~\AA\ line is
taken from the latest NIST atomic transition probability ``critical
compilation'' (Fuhr \& Wiese 2002\nocite{fuh02}).

{\em \ion{Mo}{1}:} The $gf$ value of the 3863.10~\AA\ line is taken
from Whaling \etal\ (1984)\nocite{wha84}.

{\em \ion{Ru}{1}:} The $gf$ values are taken from Wickliffe, Salih, \& 
Lawler (1994)\nocite{wic94}.

{\em \ion{Rh}{1}:} Two sources contributed the $gf$ values for this
species: Kwiatkowski \etal\ (1982)\nocite{kwi82} for 3434.89~\AA, and
Duquette \& Lawler (1985)\nocite{duq85} for 3692.36~\AA.

{\em \ion{Sn}{1}:} This species has been extensively studied in the
laboratory, and the $gf$ values used here are taken from the 
NIST compilation  (Fuhr \& Wiese 2002\nocite{fuh02}).

{\em \ion{Yb}{2}:} Theoretical $gf$ values have been computed by
Bi\'emont \etal\ (1998)\nocite{bie98}.
We renormalized these using the accurate beam-laser lifetimes of
Pinnington, Rieger, \& Kernahan (1997)\nocite{pin97}.
There are seven stable isotopes of Yb to account for, 
\iso{168,170-174,176}{Yb}, and only \iso{168}{Yb} may be neglected
due to its very small fractional contribution (0.13\%) to the
total solar Yb abundance.
The odd-numbered isotopes, \iso{171,173}{Yb} have significant hfs.
Data for isotopic wavelength shifts and hfs were taken from
Martensson-Pendrill, Gough, \& Hannaford (1994)\nocite{mar94}

{\em \ion{Lu}{2}:} The $gf$ values were taken from Quinet \etal\ 
1999\nocite{qui99}, renormalized with the recent lifetime results
of Fedchak \etal\ (2000)\nocite{fed00}.
Hyperfine structure constants from new laboratory measurements are 
reported in Appendix~B.  
These new results are in agreement with a smaller set of published 
hfs measurements (Brix \& Kopferman 1952\nocite{bri52}).
There is only one stable isotope for this element, \iso{175}Lu.
One of the transitions employed in our Lu analysis, 3397.07~\AA,
was studied in the solar spectrum by Bord, Cowley, \& Mirijanian 
(1998)\nocite{bor98}, who give an extensive discussion of the surrounding
complex atomic and molecular spectrum.

\section{New Laboratory Data for Holmium and Lutetium}

New laboratory measurements of transition probabilities and/or hfs 
constants for \ion{Ho}{2} and \ion{Lu}{2} were performed 
as part of this study of \cs22.   
Although there is an efficiency advantage in making a large set of 
measurements once an experiment is operational, the need for improved 
laboratory data on key lines took priority here.   
A more comprehensive set of measurements will be published later.

The spectrum of singly ionized holmium needs attention from laboratory 
spectroscopists. 
There is evidence that available energy levels have errors 
of several times 0.1~cm$^{-1}$ (Worm \etal\ 1990)\nocite{wor90}.  
Only a very few laser calibrated transition probabilities and hfs constants 
have been published (Worm \etal\ 1990)\nocite{wor90}.   
Although radiative lifetimes for 37 levels of \ion{Ho}{2} were measured 
recently using laser induced fluorescence (Den~Hartog \etal\ 
1999\nocite{den99}), these lifetimes have not yet been combined with
branching fractions to determine a large set of transition
probabilities.   
Incomplete knowledge of the energy levels has made it difficult to measure 
complete sets of branching ratios or branching fractions.  
We have measured branching ratios for four lines needed in this
study of \cs22.  
Our measurements are from spectra recorded using the 1.0~m Fourier 
transform spectrometer (FTS) at the U. S. National Solar Observatory.  
The analysis procedure is very similar to that used in recent work on 
\ion{La}{2} (Lawler \etal\ 2001\nocite{law01}).    
The new \ion{Ho}{2} branching fractions, listed in Table~8, 
were combined with radiative lifetimes from Den Hartog \etal\ to yield the 
\ion{Ho}{2} log~$gf$ values of Table~3.  
The strong blue/UV transitions listed in the table are the 6p-6s branches
from each upper level.  
Efforts to observe and measure the weak infrared 6p-5d branches have 
been partially successful.  
Although the infrared branches have been omitted from Table~8, their 
approximate effect in decreasing the dominant blue/UV branching 
fractions has been included.

The same FTS spectra were used to extract hfs constants and improved 
energy levels needed in this study.  
The only stable isotope of Ho has mass 165 and nuclear spin 7/2.   
We used the ground term hfs constant reported by Worm \etal\
(1990)\nocite{wor90} in our analysis of the FTS data.   
The least-square fits to partially resolved hfs patterns are most
stable if either the upper or lower level constants are fixed.   
Our uncertainties are thus limited by the accuracy of the hfs constants from 
Worm \etal\ which are included with our results in Table~9.
Our improved energy levels are accurate to $\approx$0.003~cm$^{-1}$.

A similar analysis of FTS data was used to extract the hfs constants for 
\ion{Lu}{2}, which are given in Table~10.  
The dominant stable isotope of Lu has mass 175 and nuclear spin 7/2.    
In this case it is not necessary to use any published hfs data in the 
least-square analysis of our FTS data because we could start the analysis 
on lines connected to J~=~0 levels.   
Measurements from Brix \& Kopfermann (1952)\nocite{bri52} are 
included for comparison.

\newpage

\newpage

\newpage

\newpage

\newpage
\begin{figure}
\epsscale{0.9}
\plotone{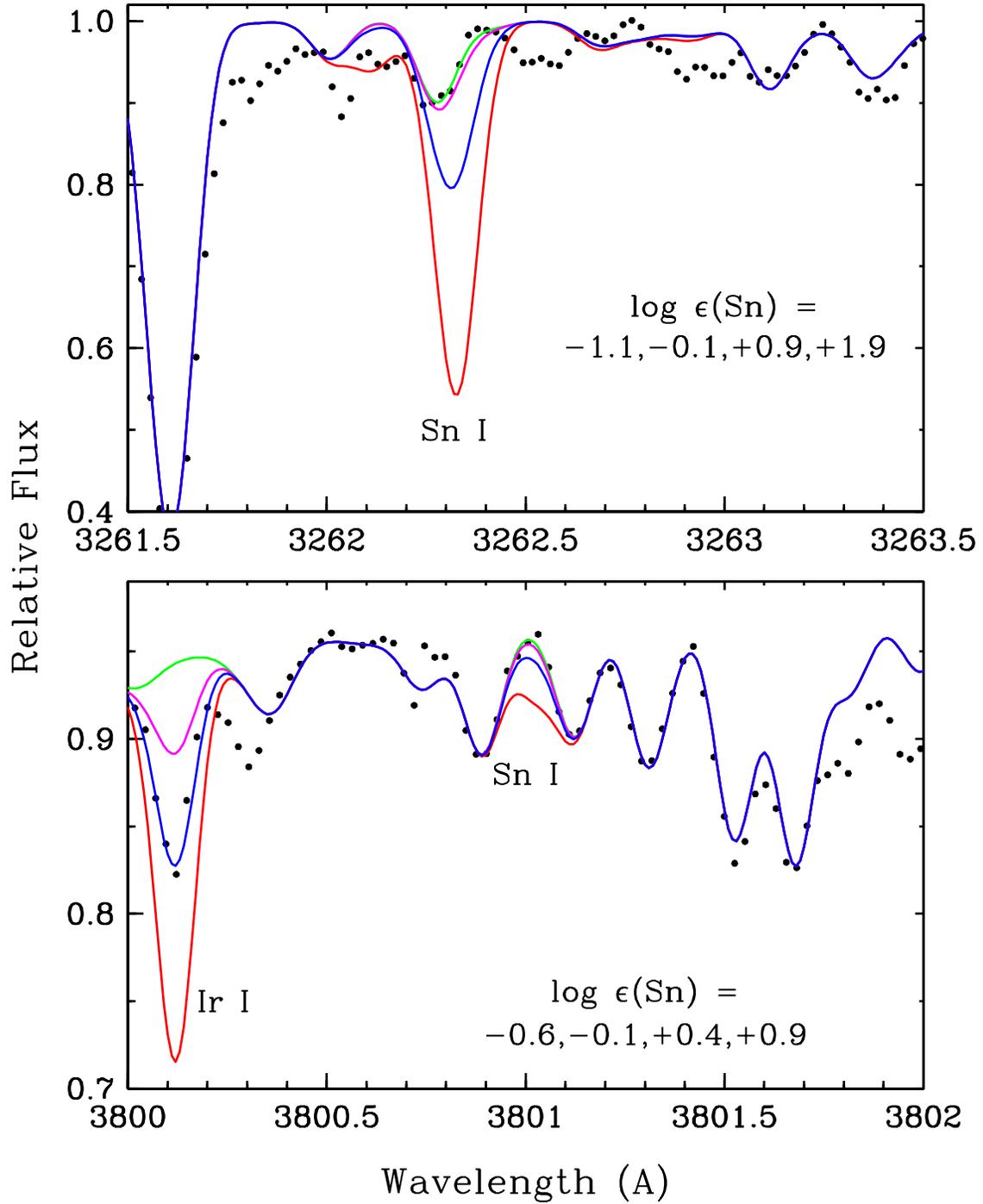}
\caption{
Observed and synthetic spectra of two \ion{Sn}{1} lines in \cs22.
The assumed Sn abundances for the synthetic spectra are given
in each figure panel.
The synthetic spectra for two smallest abundances in each case
produce nearly the same (extremely weak) \ion{Sn}{1} absorptions.
These Sn lines cannot be positively detected in \cs22.
\label{f5}}
\end{figure}

\newpage
\begin{figure}
\epsscale{1.0}
\plotone{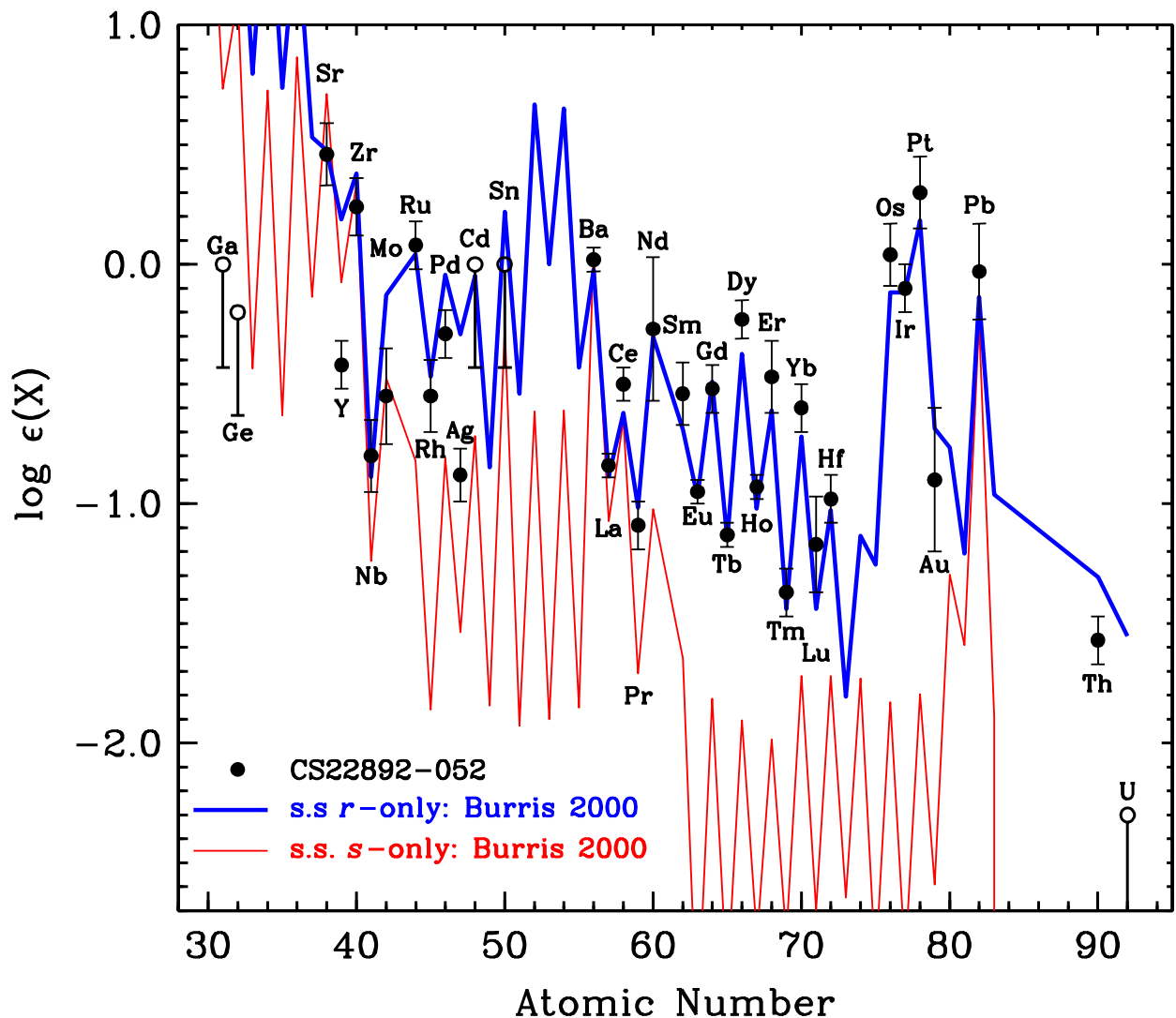}
\caption{
Observed $n$-capture abundances in \cs22\ and two scaled solar-system
abundance distributions (Burris \etal\ 2000).
Detected elements are shown as solid circles with complete error bars, and
upper limits are denoted with open circles with only the lower half
of an arbitrary-length error bar.
The solar-system $r$-process abundance set (solid line) is vertically
scaled by a single additive constant to match the observed Eu abundance,
while the $s$-process set (dashed line) is scaled to match the observed
Ba abundance.
\label{f6}}
\end{figure}

\newpage
\begin{figure}
\epsscale{1.0}
\plotone{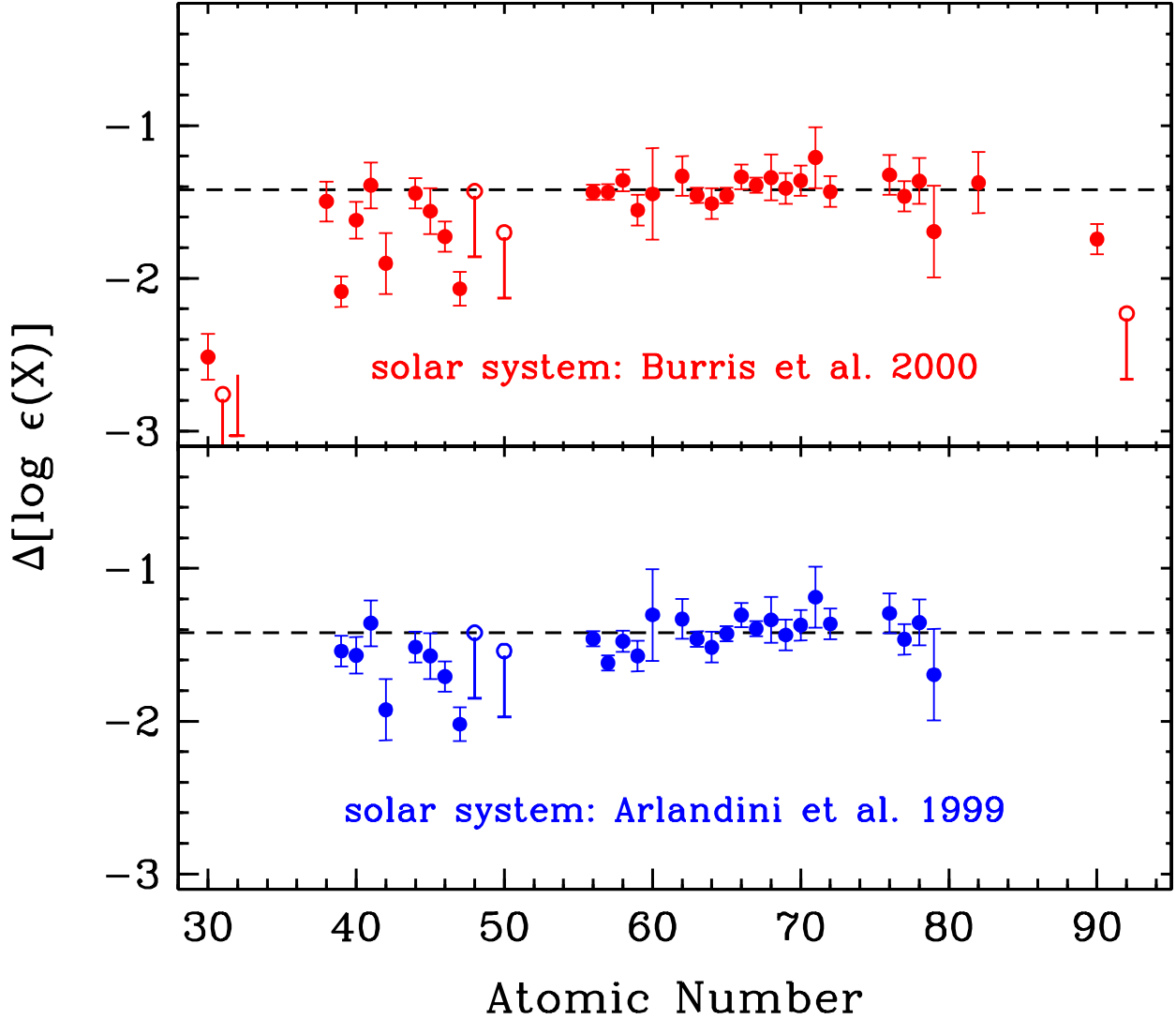}
\caption{
Differences between \cs22 abundances and two scaled solar-system abundance
distributions.
The dotted line in each panel indicates the unweighted mean difference 
for elements in the range 56~$\leq$~Z~$\leq$~79.
The symbols are as in Figure~\ref{f6}.
In the top panel the abundance differences relative to those of 
Burris \etal\ (2000; their Table~5) are displayed.  
The upper limit for Ge (Z~=~32) has no open-circle head because its upper
limit lies below the lower limit boundary of the plot.
In the bottom panel the abundance differences are relative to those 
of Arlandini \etal\ (1999).
Their solar-system $n$-capture abundances were tabulated only for
the atomic number range 39~$\leq$~Z~$\leq$~81.
\label{f7}}
\end{figure}

\newpage
\begin{figure}
\epsscale{1.0}
\plotone{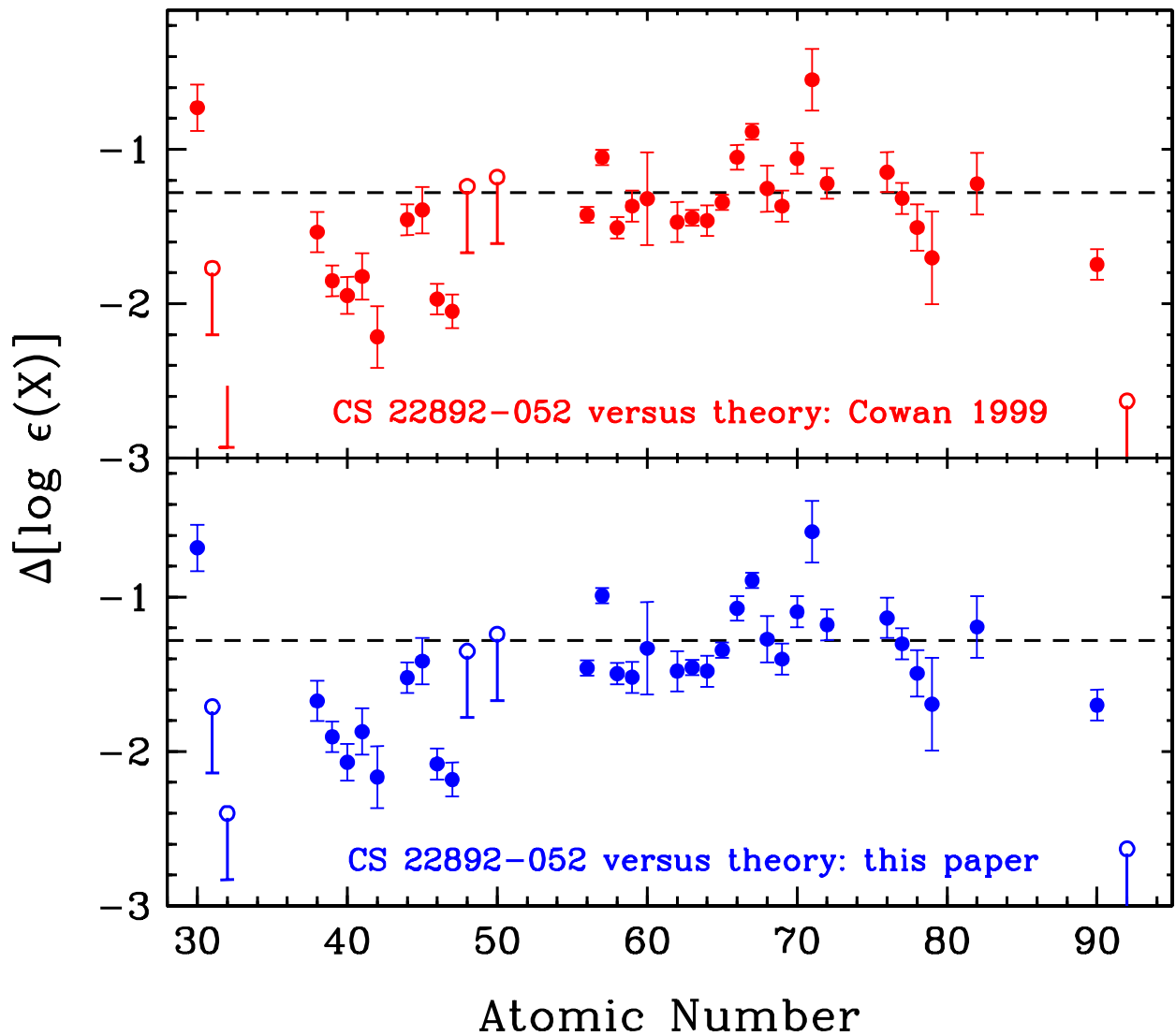}
\caption{
Differences between \cs22 abundances and two scaled $r$-process theoretical
predictions.
The symbols and lines have the same meaning as in Figure~\ref{f7}.
In the top panel the abundance differences relative to those of Cowan 
\etal\ (1999) are displayed.  
In the bottom panel the abundance differences are relative to those 
newly computed for this paper.
\label{f8}}
\end{figure}

\include{tab1}

\include{tab2}

\include{tab3}

\include{tab4}

\include{tab5}

\include{tab6}

\include{tab7}

\include{tab8}

\include{tab9}

\include{tab10}

\end{document}

%% file: tab1.tex
\label{tab1}
\tablenum{1}
\tablecolumns{3}
\tablewidth{0pt}

\begin{deluxetable}{@{\extracolsep{0.45in}}lcc}

\tabletypesize{\normalsize}
\tablecaption{Basic Data and Model Parameters}
\tablehead{
\colhead{Quantity}                         &
\colhead{Value}                            &
\colhead{Source}                           
}
\startdata
R.A. (J2000)            &  $22^h17^m01.5^s$                 & 1     \\
Dec. (J2000)            &  $-16\arcdeg 39\arcmin 26\arcsec$ & 1     \\
$l^{\rm II}$            &  41.15$\arcdeg$                   & 1     \\
$b^{\rm II}$            &  --52.8$\arcdeg$                  & 1     \\
$V$                     &  13.18                            & 2     \\
$B-V$                   &   0.78                            & 2     \\
$U-B$                   &   0.14                            & 2     \\
$V-R$                   &   0.48                            & 2     \\
$J-H$                   &   0.47                            & 3     \\
$H-K$                   &   0.11                            & 3     \\
$J-K$                   &   0.58                            & 3     \\
$V-K$                   &   2.26                            & 4     \\
$E(B-V)$                &   0.02, 0.00, 0.03                & 2,5,6 \\
$M_V$                   & --0.23, --0.16                    & 2,5   \\
Distance                &   4700, 4648 pc                   & 2,5   \\
\mbox{[Fe/H]$_{\rm C}$} & --2.92, --3.01                    & 2,5   \\
T$_{eff}$               &   4800 K                          & 7     \\
log $g$                 &   1.50                            & 7     \\
\mbox{[M/H]}            &  --2.5                            & 7     \\
v$_t$                   &   1.95 km s$^{-1}$                & 7     \\
\enddata

\tablerefs{1. SIMBAD;
           2. Beers et al. (1992);
           3. Two Micron All Sky Survey (2MASS);
           4. V from Beers et al (1992) and K from 2MASS;
           5. Beers et al. (1999);
           6. Schlegel et al. 1998;
           7. this paper.
}

\end{deluxetable}

%% file: tab2.tex
\tablenum{2}
\tablecolumns{5}
\tablewidth{0pt}

\begin{deluxetable}{lrrcr}

\tabletypesize{\footnotesize}
\tablecaption{Lines of Elements with Z $<$ 30 in CS~22892-052}
\tablehead{
\colhead{$\lambda$}                        &
\colhead{E.P.}                             &
\colhead{log $gf$}                         &
\colhead{E.W.\tablenotemark{a}}            &
\colhead{log $\epsilon$}                   \\
\colhead{\AA}                              &
\colhead{eV}                               &
\colhead{}                                 &
\colhead{m\AA}
}

\startdata
\multicolumn{5}{c}{Na I, Z = 11} \\
   3302.37 &    0.00  &--1.74   &    syn  &     +3.21 \\    
   3302.98 &    0.00  &--2.04   &    syn  &     +2.76 \\    
   5889.95 &    0.00  & +0.11   & syn(150)&     +3.11 \\    
   5895.92 &    0.00  &--0.19   & syn(122)&     +3.06 \\    
\multicolumn{5}{c}{Mg I, Z = 12} \\
   3829.36 &    2.71  &--0.21   &  138.0  &     +4.82 \\
   3832.31 &    2.71  & +0.15   &  163.0  &     +4.82 \\
\enddata

\tablenotetext{a}{An EW value in this column means that the abundance
                  was determined using this EW; the designation ``syn''
                  means that a synthetic spectrum was used; and
                  ``syn(EW)'' means that a synthetic spectrum was used,
                  but the feature was sufficiently isolated for an 
                  approximate EW (to indicate absorption strength) 
                  to be measured.}

\end{deluxetable}

%% file: tab3.tex
\tablenum{3}
\tablecolumns{5}
\tablewidth{0pt}

\begin{deluxetable}{lrrcr}

\tabletypesize{\footnotesize}
\tablecaption{Lines of $n$-Capture Elements in CS~22892-052}
\tablehead{
\colhead{$\lambda$}                        &
\colhead{E.P.}                             &
\colhead{log $gf$}                         &
\colhead{EW\tablenotemark{a}}              &
\colhead{log $\epsilon$}                   \\ 
\colhead{\AA}                              &
\colhead{eV}                               &
\colhead{}                                 &
\colhead{m\AA}                              
}

\startdata
\multicolumn{5}{c}{Ga I, Z = 31} \\
   4172.04 &    0.10  &--0.32   &    syn  &  $<$+0.0  \\    
\multicolumn{5}{c}{Ge I, Z = 32} \\   
   3039.07 &    0.88  &--0.04   &    syn  & $<$--0.2  \\    
\multicolumn{5}{c}{Sr I, Z = 38} \\
   4607.33 &    0.00  & +0.28   &    syn  &     +0.5: \\    
\multicolumn{5}{c}{Sr II, Z = 38} \\
   3464.46 &    3.04  &  0.53   &    syn  &     +0.3: \\
\enddata

\tablenotetext{a}{An EW value in this column means that the abundance
                  was determined using this EW; the designation ``syn''
                  means that a synthetic spectrum was used; and
                  ``syn(EW)'' means that a synthetic spectrum was used,
                  but the feature was sufficiently isolated for an 
                  approximate EW (to indicate absorption strength) 
                  to be measured.}

\end{deluxetable}

%% file: tab4.tex
\tablenum{4}
\tablecolumns{9}
\tablewidth{0pt}

\begin{deluxetable}{lrrrrcrrr}

\tabletypesize{\normalsize}
\tablecaption{Abundances for Lighter Elements}
\tablehead{
\colhead{Species}                          &
\colhead{Z}                                &
\colhead{log $\epsilon$}                   &
\colhead{$\sigma$}                         &
\colhead{ebar\tablenotemark{a}}            &
\colhead{N$_{lines}$}                      &
\colhead{log $\epsilon_\sun$\tablenotemark{b}}              &
\colhead{[X/H]}                            &
\colhead{[X/Fe]}                           
}
\startdata
Li I         &  3 &  +0.15 & \nodata & 0.30&   1 &\nodata &\nodata &\nodata \\
Be I         &  4 & --2.4: & 0.30    & 0.30&   2 &\nodata &\nodata &\nodata \\
CH           &  6 &  +6.30 & \tablenotemark{c} 
                                     & 0.10&   1 &   8.52 & --2.22 &  +0.88 \\
CN           &  7 &  +5.83 & \tablenotemark{c}
                                     & 0.20&   1 &   7.92 & --2.09 &  +1.01 \\
\mbox{[O I]} &  8 &  +6.45 & \nodata & 0.15&   1 &   8.83 & --2.38 &  +0.72 \\
Na I         & 11 &  +3.04 & 0.19    & 0.19&   4 &   6.33 & --3.29 & --0.19 \\
Mg I         & 12 &  +4.78 & 0.08    & 0.08&   7 &   7.58 & --2.87 &  +0.30 \\
Al I         & 13 &  +2.79 & \nodata & 0.15&   1 &   6.47 & --3.68 & --0.58 \\
\enddata

\tablenotetext{a}{Adopted error bar; see text for further discussion.}

\tablenotetext{b}{Grevesse \& Sauval (1998) solar photospheric abundances.}

\tablenotetext{c}{Abundance determined from a synthetic spectrum computation
                  involving numerous molecular band lines.}

\end{deluxetable}

%% file: tab5.tex
\tablenum{5}
\tablecolumns{9}
\tablewidth{0pt}

\begin{deluxetable}{lrrrrcrrr}

\tabletypesize{\footnotesize}
\tablecaption{Abundances for Neutron-Capture Elements}
\tablehead{
\colhead{Species}                          &
\colhead{Z}                                &
\colhead{log $\epsilon$}                   &
\colhead{$\sigma$}                         &
\colhead{ebar\tablenotemark{a}}            &
\colhead{N$_{lines}$}                      &
\colhead{log $\epsilon_\sun$\tablenotemark{b}}              &
\colhead{[X/H]}                            &
\colhead{[X/Fe]}                           
}
\startdata
Ga I       & 31 &  $<$0.0& \nodata&\nodata&   1 &  2.88 &$<$--2.9 & $<$+0.2 \\
Ge I (HST) & 32 &$<$--0.2& \nodata&\nodata&   1 &  3.41 &$<$--3.6 &$<$--0.5 \\
Sr I       & 38 &   +0.50& \nodata&  0.15 &   1 &  2.97 &  --2.47 &   +0.63 \\
Sr II      & 38 &   +0.45& 0.13   &  0.13 &   4 &  2.97 &  --2.52 &   +0.58 \\
Y  II      & 39 &  --0.42& 0.10   &  0.10 &  15 &  2.24 &  --2.66 &   +0.44 \\
Zr II      & 40 &   +0.23& 0.12   &  0.12 &  28 &  2.60 &  --2.37 &   +0.73 \\
Zr II (HST)& 40 &   +0.32& 0.18   &  0.18 &   3 &  2.60 &  --2.28 &   +0.82 \\
Nb II      & 41 &  --0.80& \nodata&  0.15 &   1 &  1.42 &  --2.22 &   +0.88 \\
\enddata

\tablenotetext{a}{Adopted error bar; see text for further discussion}

\tablenotetext{b}{Grevesse \& Sauval (1998) solar photospheric abundances,
except for the poorly-determined values for Ag, Tb, Ho, Tm, Au, and U; for
these elements the meteoritic abundances are listed.}

\end{deluxetable}

%% file: tab6.tex
\tablenum{6}
\tablecolumns{6}
\tablewidth{0pt}

\begin{deluxetable}{lccccccc}

\tabletypesize{\normalsize}
\tablecaption{Model Fits to the Data}
\tablehead{
\colhead{Model}                   &
\colhead{Mean Difference}                         &
\colhead{$\sigma$}                      &
\colhead{Model}              &
\colhead{Mean Difference}                            &
\colhead{$\sigma$}                          \\ 
\colhead{}                   &
\colhead{(light)}                         &
\colhead{(light)}                      &
\colhead{}              &
\colhead{(heavy)}                            &
\colhead{(heavy)}
}
\startdata
Burris et al.\tablenotemark{a}      &  -1.70 $\pm$ 0.09  & 0.26  & 
Burris et al.\tablenotemark{b}      & -1.42  $\pm$ 0.02 &  0.10    \\
Arlandini et al.\tablenotemark{c}  &  -1.65  $\pm$ 0.08  & 0.22   & 
Arlandini et al.\tablenotemark{b}  &  -1.42  $\pm$ 0.03  & 0.12   \\
Old Theory\tablenotemark{a}      &  -1.80  $\pm$ 0.09    & 0.28  & 
Old Theory\tablenotemark{b}     &   -1.28  $\pm$ 0.06    & 0.26   \\
New Theory\tablenotemark{a}    &   -1.88   $\pm$ 0.09  &  0.28  &
New Theory\tablenotemark{b}     &   -1.28   $\pm$ 0.06  &  0.26  \\
\enddata

\tablenotetext{a}{Z=38--55}
\tablenotetext{b}{Z=56--79}
\tablenotetext{c}{Z=39--55}

\end{deluxetable}

%% file: tab7.tex
\tablenum{7}
\tablecolumns{6}
\tablewidth{0pt}

\begin{deluxetable}{c@{\extracolsep{0.10in}}ccccc}
\tablecaption{Chronometric Age Estimates for CS~22892--052}
\tablehead{
\colhead{Chronometer Pair}                 &
\colhead{Predicted\tablenotemark{a}}                        &
\colhead{Observed}                         &
\colhead{Average Age}                              &
\colhead{Solar\tablenotemark{b}}           &
\colhead{Lower Limit}                      \\
                                           &
                                           &
                                           &
\colhead{(Gyr)}                            &
                                           &
\colhead{(Gyr)}
}
\startdata
Th/Eu  &    0.45    &   0.24    &   12.8 & 0.4615  &   13.2 \\
Th/Ir  &    0.0876   &   0.0339  &   19.2 & 0.0646  &  13.1 \\
Th/Pt  &    0.0226   &   0.0135   &   10.5 & 0.0323  &  17.7 \\
       &             &            &        &         &       \\
Th/U  &  1.45--1.805  &   $\ge$ 5.4    & $\ge$  10.4 & ---    &  --- \\
\enddata

\tablenotetext{a}{Average from Cowan et al. (1999) and the current paper.}

\tablenotetext{b}{From Burris et al. (2000).}

\end{deluxetable}

%% file: tab8.tex
\tablenum{8}
\tablecolumns{7}
\tablewidth{0pt}

\begin{deluxetable}{@{\extracolsep{0.15in}}lrrcrcr}

\tabletypesize{\normalsize}
\tablecaption{Branching Fractions for \ion{Ho}{2}}
\tablehead{
\colhead{Transition}                       &
\colhead{$\lambda$}                        &
\colhead{E$_{\rm upper}$}                  &
\colhead{J$_{\rm upper}$}                  &
\colhead{E$_{\rm lower}$}                  &
\colhead{J$_{\rm lower}$}                  &
\colhead{Branching}                        \\
\colhead{cm$^{-1}$}                        &
\colhead{nm}                               &
\colhead{cm$^{-1}$}                        &
\colhead{}                                 &
\colhead{cm$^{-1}$}                        &
\colhead{}                                 &
\colhead{Ratio}                                 
}
\startdata
26234.36  &  381.071  &  26234.36  &  7  &    0.00  &  8  &  0.805$\pm$3\% \\ 
25596.61  &  390.566  &  26234.36  &  7  &  637.75  &  7  &  0.145$\pm$3\% \\
24712.16  &  404.545  &  24712.16  &  7  &    0.00  &  8  &  0.845$\pm$3\% \\
24074.41  &  415.262  &  24712.16  &  7  &  637.75  &  7  &  0.105$\pm$3\% \\
\enddata

\end{deluxetable}

%% file: tab9.tex
\tablenum{9}
\tablecolumns{6}
\tablewidth{0pt}

\begin{deluxetable}{@{\extracolsep{0.20in}}cccrrr}

\tabletypesize{\normalsize}
\tablecaption{Improved Energy Levels and Hyperfine Structure Constants for 
              \ion{Ho}{2}\tablenotemark{a}}
\tablehead{
\colhead{Configuration}                    &
\colhead{Term}                             &
\colhead{J}                                &
\colhead{E}                                &
\colhead{A}                                &
\colhead{B}                                \\
\colhead{}                                 &
\colhead{}                                 &
\colhead{}                                 &
\colhead{cm$^{-1}$}                        &
\colhead{mK\tablenotemark{b}}              &
\colhead{mK\tablenotemark{b}}               
}
\startdata
${\rm 4f^{11}(^{4}I_{15/2})6s_{1/2}}$ & (15/2,1/2)$^{\rm o}$ &  8 &     0.000 &
+52.64$\pm$0.07\tablenotemark{c} &  --52.6$\pm$3.3\tablenotemark{c} \\
${\rm 4f^{11}(^{4}I_{15/2})6s_{1/2}}$ & (15/2,1/2)$^{\rm o}$ &  7 &   637.754 &
 +0.83$\pm$0.07\tablenotemark{c} &  --56.4$\pm$3.5\tablenotemark{c} \\
                             \nodata  &           \nodata    &  7 & 24712.163 &
+42.15$\pm$0.05                  &    +0.2$\pm$2.0                  \\
${\rm 4f^{11}(^{4}I_{15/2})6p_{1/2}}$ & (15/2,1/2)           &  7 & 26234.363 &
+41.60$\pm$0.05                  &   +21.6$\pm$2.0                  \\
\enddata

\tablenotetext{a}{\hspace*{0.10in}$^{165}$Ho, I~=~7/2}

\tablenotetext{b}{\hspace*{0.10in}1.0~mK = 10$^{-3}$~cm$^{-1}$}

\tablenotetext{c}{\hspace*{0.10in}Worm et al. (1990)}

\end{deluxetable}

%% file: tab10.tex
\tablenum{10}
\tablecolumns{8}
\tablewidth{0pt}

\begin{deluxetable}{cccrccrr}

\tabletypesize{\normalsize}
\tablecaption{Hyperfine Structure Constants for 
              \ion{Lu}{2}\tablenotemark{a}}
\tablehead{
\colhead{Configuration}                    &
\colhead{Term}                             &
\colhead{J}                                &
\colhead{E}                                &
\colhead{A}                                &
\colhead{B}                                & 
\colhead{A\tablenotemark{b}}               &
\colhead{B\tablenotemark{b}}               \\
\colhead{}                                 &
\colhead{}                                 &
\colhead{}                                 &
\colhead{cm$^{-1}$}                        &
\colhead{mK\tablenotemark{c}}              & 
\colhead{mK\tablenotemark{c}}              & 
\colhead{mK\tablenotemark{c}}              & 
\colhead{mK\tablenotemark{c}}               
}
\startdata

5d6s & $^{\rm 3}$D           & 1 & 11796.24 &  --68.32$\pm$0.07 &  
                       +29.3$\pm$0.7 & --68.2           &  +29.0         \\ 
5d6s & $^{\rm 3}$D           & 2 & 12435.32 &   +64.39$\pm$0.10 &  
                       +45.8$\pm$1.0 &  +64.5           &  +46.4         \\
5d6s & $^{\rm 3}$D           & 3 & 14199.08 &   +57.06$\pm$0.20 &  
                       +79.2$\pm$2.0 &  +56.8           &  +76.8         \\
5d6s & $^{\rm 1}$D           & 2 & 17332.58 &  --25.71$\pm$0.20 &  
                       +74.7$\pm$2.0 & \nodata          &  \nodata       \\
5d6p & $^{\rm 3}$P$^{\rm o}$ & 1 & 28503.16 &  +165.38$\pm$0.07 & 
                      --61.1$\pm$0.7 & +165.4           & --62.5         \\
5d6p & $^{\rm 3}$P$^{\rm o}$ & 2 & 32452.26 &   +95.60$\pm$0.10 & 
                      +115.4$\pm$1.5 &  +95.9           & +113.6         \\
5d6p & $^{\rm 3}$F$^{\rm o}$ & 2 & 41224.96 &   +26.95$\pm$0.10 &  
                       +63.9$\pm$1.0 & \nodata          &  \nodata       \\
5d6p & $^{\rm 3}$F$^{\rm o}$ & 3 & 44918.68 &   +10.90$\pm$0.20 & 
                      +105.2$\pm$2.0 & \nodata          &  \nodata       \\
\enddata

\tablenotetext{a}{\hspace*{0.10in}$^{175}$Lu, I~=~7/2} 

\tablenotetext{b}{\hspace*{0.10in}Brix \& Kopfermann (1952)}

\tablenotetext{c}{\hspace*{0.10in}1.0~mK = 10$^{-3}$~cm$^{-1}$}

\end{deluxetable}